\documentclass[letterpaper,10pt,twocolumn]{article}
\usepackage[pdftex]{graphicx}
\usepackage{usenix}

\usepackage{times,color}
\usepackage{amsmath}
\usepackage{paralist}           %
\usepackage{balance}            %

\usepackage{xcolor}
\definecolor{greenLight}{HTML}{C0DF9A}
\definecolor{greenLogo}{HTML}{96CA50}
\definecolor{greenDark}{HTML}{668B37}
\definecolor{accentLight}{HTML}{D5AEAA}
\definecolor{accent}{HTML}{AC5D55}
\definecolor{accentDark}{HTML}{820C00}
\definecolor{emphasisLight}{HTML}{FFE0A2}
\definecolor{emphasis}{HTML}{FFCA5F}
\definecolor{emphasisDark}{HTML}{B78B41}
\definecolor{neutralLighter}{HTML}{D6DBE4}
\definecolor{neutralLight}{HTML}{8897AE}
\definecolor{neutral}{HTML}{385378}
\definecolor{neutralDark}{HTML}{263852}

   \usepackage[square,sort&compress, authoryear]{natbib}
\newenvironment{acks}{\section*{Acknowledgements}}{}

\renewcommand{\cite}[1]{\errmessage{Don't use cite, you probably want citep}}

\usepackage{verbatim}           %
\newcommand{\code}[1]{\texttt{#1}}

\usepackage{xspace}

\usepackage{subcaption}

\usepackage[pdftex]{graphicx}
\setkeys{Gin}{keepaspectratio=true,clip=true,draft=false}
\graphicspath{{./imgs/}}

\usepackage[pdftex]{hyperref}
\hypersetup{nolinks=false}

  \newcommand{\lions}{\mbox{LionsOS}\xspace}

\newcommand{\ISC}{I\textsuperscript{2}C\xspace}

\usepackage{multirow}

\begin{document}\sloppy

  \renewcommand{\sectionautorefname}{Section}
  \renewcommand{\subsectionautorefname}{Section}
  \renewcommand{\subsubsectionautorefname}{Section}
  \renewcommand{\appendixautorefname}{Appendix}
  \renewcommand{\Hfootnoteautorefname}{Footnote}
  \newcommand{\Htextbf}[1]{\textbf{\hyperpage{#1}}}

\renewcommand{\topfraction}{0.4}
\renewcommand{\bottomfraction}{0.4}

\title{Fast, Secure, Adaptable: \lions Design, Implementation and Performance}
  \author{Gernot Heiser, Ivan Velickovic, Peter Chubb\\
    Alwin Joshy, Anuraag Ganesh, Bill Nguyen, Cheng Li, Courtney
    Darville, Guangtao Zhu, \\
    James Archer, Jingyao Zhou, Krishnan
    Winter, Lucy Parker, Szymon Duchniewicz, Tianyi Bai \\
    UNSW\\
    Sydney, Australia\\
    \href{mailto:gernot@unsw.edu.au}{\{gernot, peter.chubb, i.velickovic\}@unsw.edu.au}
  }

\maketitle
\subsection*{Abstract}

We present \lions, an operating system for security- and
safety-critical embedded systems. \lions is based on the formally
verified seL4 microkernel and designed with verification in mind. It
uses a static architecture and
features a highly modular design driven by strict separation of
concerns and a focus on simplicity.  We demonstrate that \lions
achieves excellent performance on system-call intensive workloads.

\section{Introduction}\label{s:intro}

Safety- and security-critical systems, such as aircraft, autonomous
cars, medical devices, industrial control or defence systems, require a highly
dependable operating system (OS).  The complexity (and code size) of
these embedded/cyber-physical systems keeps growing, and it is unavoidable to have highly
critical (and, presumably, highly assured) functionality co-exist with
less critical (and less trustworthy) functionality. These systems are
therefore \emph{mixed criticality systems} (MCS), where the correct
operation of a critical component must not depend on the correctness
of a less critical component \citep{Barhorst_BBHPSSSSU_09, Burns_Davis_17}.

Traditional MCS are concerned with (spatial and temporal)
\emph{integrity and availability}. In modern systems the mixed-criticality requirement
must also extend to \emph{confidentiality}: Many
cyberphysical systems, such as drones or medical devices, process
sensitive data that must be protected.

The core requirement such critical systems impose on the OS is therefore strong
temporal and spatial isolation.  As such, the seL4
microkernel \citep{seL4:URL} seems a perfect foundation: seL4
has undergone extensive formal verification,
including proofs of confidentiality and integrity enforcement, proof
of implementation correctness and proofs that the binary has the same
semantics as the verified C code, taking the compiler out of the
trusted computing base (TCB) \citep{Klein_AEMSKH_14}. It is the only
formally-verified kernel that uses fine-grained access control through capabilities
\citep{Dennis_VanHorn_66}. There
exists an MCS version of seL4 that provides the temporal
isolation properties required by real-time
systems \citep{Lyons_MAH_18}, verification of the MCS variant is
currently in progress \citep{seL4:roadmap}.

These features have resulted in some real-world deployments, including
autonomous military aircraft \citep{Cofer_GBWPFPKKAHS_18} and, more
recently, commercial electric cars \citep{Qu_24:sel4s}.  Yet more than a decade
after seL4's verification was completed, it is not widely deployed.

The core reason behind this slow uptake seems to be the low-level nature of
seL4, even lower than for most other microkernels.
For example, seL4 makes management of all of a
system's physical memory (including the kernel's) a user-level
responsibility \citep{Elkaduwe_DE_08}; while this is a core enabler of
reasoning about isolation, managing memory requires data structures
(e.g.\ page tables) which require memory -- a foot gun for developers.

seL4 can be said to be the ``assembly language of
operating systems''.
To build functional, performant systems on seL4 requires deep expertise -- the
kernel is largely unapproachable by industrial developers of critical
systems, resulting in a number of seL4-based deployment projects
apparently abandoned a few years after they started.

In short, to benefit from seL4's provable isolation enforcement,
developers of critical systems need an actual \emph{operating system}, providing
appropriate abstractions, such as processes, files and network
connections, while retaining (and extending) the isolation guarantees
provided by seL4.  We furthermore posit that such a system must be
highly modular, in order to make best use of seL4-provided isolation
for  minimising the impact of faults, simplify identification and
elimination of bugs, and enable end-to-end formal verification
(eventually -- for now we leave verification out of scope but aim for a
verification-friendly design.)

Fine-grained modularity results in many context switches and IPC
operations; this overhead has traditionally been considered
the Achilles heel of microkernels \citep{Bershad_92}. Even very recent
microkernel-based work co-locates services in larger modules and
migrates functionality back into the kernel in order to achieve
performance competitive with the monolithic Linux system
\citep{Chen_JWLLLWHLYWYPX_24}.

While such performance challenges may be real for a general-purpose OS
system (eg.\ for smartphones), we \textbf{aim to achieve a performant yet
highly-modular OS} for the embedded/cyberphysical space.

Our \textbf{contribution} is to demonstrate
\begin{itemize}
\item \lions, the
  first highly-modular microkernel-based OS that achieves
  performance at par or better than traditional monolithic
  designs;
\item which is enabled by a simplicity-oriented design
  employing \emph{use-case-specific policies}.
\end{itemize}

We present the \emph{principles} of our approach
(\autoref{s:princip}), discuss the resulting design
(\autoref{s:design}) and its implementation focussed on simplicity
(\autoref{s:impl}). We evaluate \lions performance against Linux and
microkernel-based OSes and find that on context-switch intensive
loads, which can be expected to show the highest modularity-imposed
overheads, \lions outperforms all systems we compare against (\autoref{s:eval}).

\section{Background and Related Work}\label{s:backg}

\subsection{Verification and scalability}\label{s:b-scale}

The formal verification of seL4 demonstrated that it is possible to
prove the implementation correctness of real-world systems of
considerable complexity. However, the cost was high: about 12 person
years of non-recurring engineering for 8,500 source lines of code
(SLOC), and an estimated cost of
\$350/SLOC \citep{Klein_EHACDEEKNSTW_09}. While potentially justified
for a stable, foundational piece of infrastructure, this cost is too
high for most systems, especially a complete OS that will be significantly
larger than seL4.

The seL4 project delivered another key insight: Verification effort
scales with the square of the specification
size \citep{Matichuk_MAJKS_15}. This implies that there could be a
large scalability benefit from keeping things small and simple, i.e.\
a highly modular design, where each component has a narrow interfaces and simple
specification, and where module boundaries are enforced by seL4, making it
possible to verify modules independently of each other.

While seL4 used labour-intensive interactive theorem proving, recent
work increasingly uses automated theorem proving
techniques \citep{Sigurbjarnarson_BTW_16, Zaostrovnykh_PPAC_17,
  Nelson_SZJBTW_17, Nelson_BGBTW_19, Zaostrovnykh_PIRPAC_19,
  Narayanan_HDALZB_20, Chen_LMNB_23, Paturel_SH_23,
  Cebeci_ZZCC_24}. These automated techniques are in essence based
on (symbolic) state-space exploration with the help of heuristics to deal with
combinatorial explosion. Yet they have severe limitations in the
complexity of the code they can tackle, and will generally work best
on simple and small modules.

\subsection{Modularity in operating systems}\label{s:b-modul}

The idea of a modular OS with hardware-enforced module boundaries  is
old, going back to the original
microkernel (before the term was coined), Brinch Hansen's Nucleus
\citep{Brinch_Hansen_70}. The approach was popularised by Mach
\citep{Rashid_JOSBFGJ_89} and taken up by other microkernel systems of
the time, such as Chorus \citep{Rozier_AABGGHKLLN_88} and QNX \citep{Hildebrand_92}.

These systems were plagued by poor performance, and
functionality moved back into the kernel \citep{Welch_91}. This did not
prevent expensive debacles, such as IBM's ill-fated, Mach-based
Workplace OS \citep{Fleisch_CT_98}.

Almost all microkernel-based OSes exhibited course-grained modularity,
typically at the level of major subsystems such as file service,
networking and process management ~\citep{Rawson_97, Whitaker_SG_02a, Hartig_HFHLMP_05,
  Herder_BGHT_06, Qubes:arch}, making them too large for
verification. Nevertheless, even the most recent work argues that the
cost of crossing module boundaries is too high, resulting in a
drastically-expanded kernel of 90\,kSLOC (vs.\ seL4's 10\,kSLOC) and
co-locating services into even more course-grained isolation domains
\citep{Chen_JWLLLWHLYWYPX_24}.

The Flux OSKit \citep{Ford_BBLLS_97} was an early design featuring a
more fine-grained design. Performance comparison to Linux and FreeBSD
showed a 13\% degradation in networking throughput and a 45\% increase
in latency. SawMill \citep{Gefflaut_JPLEUTDR_00} was an ambitious project aiming to break up Linux
into components isolated by a microkernel; file-system benchmarks
showed a throughput degradation of about 18\%. No CPU load values are
reported for any of these systems, but the degradation in achieved
throughput indicates a significant increase in per-packet processing cost.

Genode (formerly Bastei) \citep{Feske_Helmuth_07, Feske:genode}
features a modular design, explicitly prioritising assurance
over performance (we could not find any published
performance data, although we evaluate against it in \autoref{s:other_os}).
Its implementation in C++ will prevent a complete
formal verification for the foreseeable future. THINK
\citep{Fassino_SLM_02} is a component system for building kernels
rather than an OS with isolated components. Systems like TinyOS \citep{Levis_05},
Tock \citep{Levy_CGGPDL_17} and Tinkertoy \citep{Wang_Seltzer_22} are
for microcontrollers without memory protection and as such unsuitable
for MCS.

An alternative to enforcing modularity by address-space isolation is
enforcing it through the programming language, as pioneered by SPIN
\citep{Bershad_SPSFBCE_95}, and later adopted by Singularity
\citep{Fahndrich_AHHHRL_06} and RedLeaf
\citep{Narayanan_HDALZB_20}. These systems generally exhibit lower
performance than mainstream OSes. Furthermore, as their security
relies on type-safety enforcement by the programming language, they
require the whole OS to be implemented in that language. Inevitably
this requires unsafe escapes for dealing with hardware. More
importantly, this rules out re-using code from mainstream
OSes.

Writing all device drivers from scratch is generally infeasible. We
therefore ignore language-based isolation approaches and instead
focus on modularity enforced by address-space isolation.

What is common to these earlier systems is their complexity, among
others driven by the desire for code or API re-use leading to design
compromises.  \textbf{Our core take-away} is that to make modular OSes
work, a clean, principled from-scratch approach is needed. Re-use,
while desirable, should be subject to clean design principles, rather
than compromising them.

\section{\lions Design Principles}\label{s:princip}

Given this experience, what makes us think we can meet
our performance aim from \autoref{s:intro} with a modular system?

We observe that a commonality of these earlier systems is a
significant complexity in design and implementation. We posit that the
key to meeting aim is a strict application of the time-honoured \emph{KISS
Principle} \citep{KISS:wp}.  Following this high-level principle, we aim
for a highly modular design which incorporates the following secondary
principles:
\begin{description}
\item[Strict separation of concerns:] Each module has one and only one
  purpose (as far as feasible).  Furthermore, a particular concern
  (e.g.\ the network traffic-shaping policy) should be
  fully contained in a single module.
\item[Least privilege:] Each module only has the access rights it
  needs, not more.  While not a consequence of KISS, this
  time-honoured security principle of
  \citet{Saltzer_Schroeder_75} simplifies reasoning about
  security and safety.
\item[Design for verification:] Module interfaces are narrow and module
  implementations simple, to make verification scalable (\autoref{s:b-scale}).
\item[Use-case specific policies:] Likely the most
  controversial, this principle calls for tailoring each (resource) policy to
  the system's particular use case, in order to simplify the policy
  implementation.
\end{description}

It seems clear that adhering to these principles, which we summarise
as ``radical simplicity'', will give us the best chance to produce a
highly dependable system and maximize the chances of formally proving
its correctness and security.

Simplicity is aided by restricting our target domain to
embedded systems.  While aiming for generality within that domain --
which specifically includes cyber-physical systems such as autonomous aircraft
and cars, some of which are
quite complex and demanding -- we do not (yet) aim to support more
general-purpose systems, such as cloud servers, smartphones and
certainly not laptops.

The common thread of the embedded domain is that it can be served with
a \emph{static system architecture}, i.e.\ a set of components that is
known at configuration time.   This does not mean that the
system is fully static -- it can support late loading of
components, dynamic component updates, and place-holders for
components that are loaded with programs not known at build
time.  Dependable embedded systems generally cannot over-commit
resources, which is what makes the static architecture work. \emph{We are
yet to see a realistic use case in the embedded space that cannot be
addressed with a static architecture.} Note that the static system
architecture does not prevent a subsystem, such as a virtual machine
(VM) from managing a subset of resources dynamically.

The inherent constraints of embedded systems are also in line with the
principle of use-case specific policies.  A computer system generally
has \emph{two classes of policies: security and resource policies}.  In the
embedded space, the security policy is generally defined by the use
case, and will only change in the context of a significant
re-configuration of the system (accompanied with a major software
upgrade).

The embedded system's defined set of resources implies that, at
least for the critical systems we are targeting, the designer has (or
should have) a clear idea of how they should be managed. The more
tailored the policy is to the use case, the simpler its
implementation, and the easier it is to assure (formally or informally)
that it matches requirements.

This principle of use-case specific policies is arguably the clearest
departure from conventional approaches.  OSes tend to be designed to
adapt to changing application scenarios with no or minimal code
changes.  This naturally leads to a (conscious or not) desire
to provide \emph{universal policies}. Of course, no policy is truly
universal, and sooner or later it will encounter a use case where the
existing policy behaves pathologically, resulting in attempts to
generalise it.  This approach is a massive driver of complexity: For
example, the Linux scheduler contains five scheduling classes, each of
which has one or more per-thread tuning parameters; the scheduler has
grown from around 11,kSLOC in 2011 (version 3.0) to over 30\,kSLOC
today (version 6.12).
Furthermore, the approach frequently
leads to optimising particular ``hot'' use cases at the expense of
overall performance \citep{Ren_RCVSY_19}.

Use-case specific policies represent the opposite approach: Each
policy is highly specialised for the use case, and the system achieves use-case
diversity not by generalising policies, but by re-writing them as needed. Of
course, this can only work if the policies are simple enough to
implement.

Our underlying argument is that by taking a radical approach to
simplicity and use-case specificity, policies \emph{do} become
simple.  Moreover, for most resources there is a small to moderate set
of policies that can be pre-supplied, letting the system designer
chose from an existing set (or trivially adapting an existing one to
the use case).

An illustrative example is network traffic shaping: If
multiple clients of a network interface overload
the interface, there are only a small number of obvious policies to
choose from: Clients may be given a priority, their bandwidth may be
limited,  they may be served round-robin, or certain protocols might
be prioritised.
In an embedded system, it
is usually obvious which one is appropriate, and this is unlikely to
change for the lifetime of the system.

Taken together, our principles lead to an OS becoming something akin to a
Lego\textsuperscript\textregistered\ set: The OS is built from different kinds of
components (brick shapes).  Each component comes in multiple,
functionally compatible versions (brick colours), and the choice of
version (colour) can be made largely independent of the rest of the system.

An obvious concern is how the fine-granular modularity
affects our performance aim, as modularity necessarily leads to high
context-switching rates.  Given there is a cost to every context switch
(depending on the architecture, 400--600 cycles for
seL4 \citep{sel4:perf}), this has the potential to make the system
slow \citep{Chen_JWLLLWHLYWYPX_24}. We will examine the performance impact in \autoref{s:eval}.

\section{\lions Design}\label{s:design}

\lions is based on seL4, and uses its address-space, thread and IPC abstractions.
Like most OSes, it consists mostly of I/O subsystems (device
drivers, protocol stacks, file systems) and resource management. The
latter part is particularly small in \lions because of its static
architecture, which reduces resource management to some simple
policy modules that are not only use-case specific (as per our
principle) but also local in their nature (e.g.\ shaping traffic of a
network interface).

Some global resource management is required,
such as core management (off- and on-lining processor cores based on
overall CPU load). This is presently work in progress. %

\subsection{Devices}

We demonstrate how our principle of strict separation of
concerns applies to I/O, using networking, specifically Ethernet, as
a case study. We briefly discuss other device classes in \autoref{s:dev-oth}.

\subsubsection{Device drivers}\label{s:des-dd}

Applying separation of concerns, we reduce the
purpose of a device driver to \emph{translate between a hardware-specific
  \underline{device interface} and a hardware-independent \underline{device class
    interface}}: an Ethernet driver does no more than abstracting the specific NIC
as a generic Ethernet device.

Unsurprisingly, the Ethernet device-class interface (i.e.\ the
driver's \emph{OS interface}) looks similar to that of an actual
Ethernet network interface controller (NIC), with some differences
that help simplify its
use. NICs typically use ring buffers in DMA memory to pass references
to DMA buffers from and to the driver; each ring buffer entry contains
a pointer to a buffer in the DMA region, together with some meta-data
(indicating whether a buffer contains valid data). A NIC usually
references two such ring buffers, one for transmit (Tx) and one for
receive (Rx) data.

The driver's OS interface uses buffer queues similar to these
hardware-specified ring buffers, but to simplify use we use separate queues for buffers
containing valid data and those that do not. This means that the software
side of the driver has four queues:
\begin{description}
\item[transmit available (TxA):] references buffers with valid data provided by the OS
  for transmission;
\item[transmit free (TxF):] references buffers returned by the NIC to the OS
  for re-use;
\item[receive available (RxA):] references buffers filled with data by
  the NIC to be consumed by the OS;
\item[receive free (RxF):] references buffers provided by the OS to
  the NIC to receive data from the network.
\end{description}

These queues are allocated in \emph{driver metadata regions}, for
Ethernet there is one for Tx data (containing the TxA and TxF queues)
and a separate one for Rx data (RxA and RxF queues). These are ``normal''
memory invisible to the device (i.e.\ not accessed by DMA). For
Ethernet we can keep the Tx and Rx regions separate.

\begin{figure}[t]
  \centering
  \includegraphics[width=\linewidth]{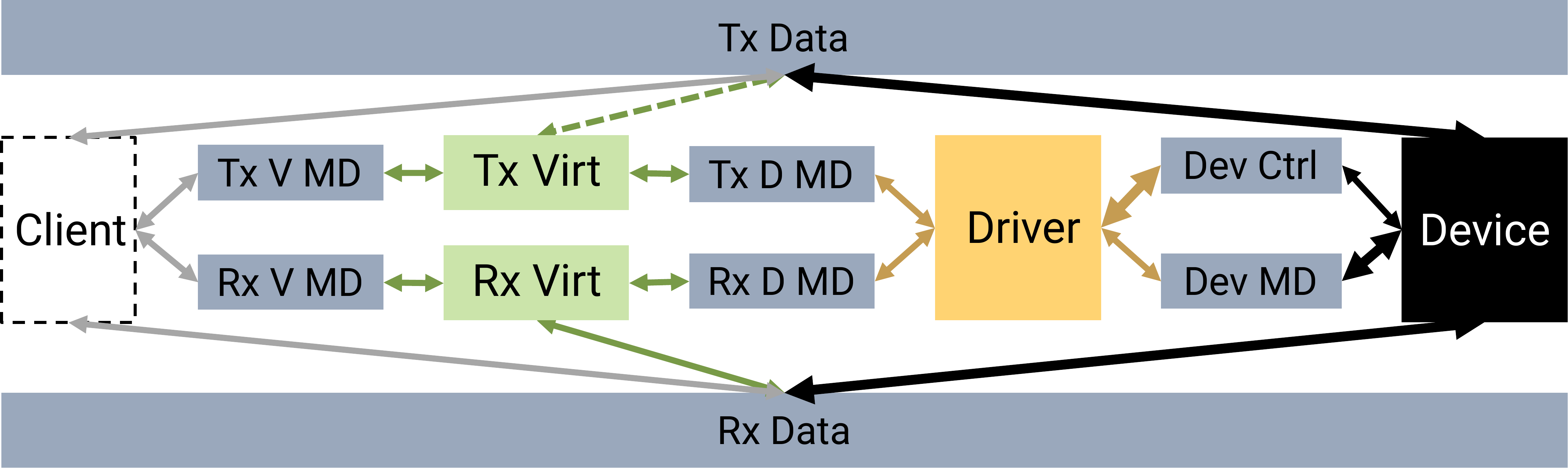}
  \caption{Memory regions for Ethernet: device control (Dev\:Ctl),
    device metadata (Dev\:MD), transmit and receive data (Tx\:Data,
    Rx\:Data), driver metadata (Tx\:D\:MD, Rx\:D\:MD) and virtualiser
    metadata (Tx\:V\:MD, Rx\:V\:MD). Arrows indicate access to regions by
    components, thick black arrows indicate DMA by the device, the
    thick coloured arrow indicates uncached access by the driver. The
    TxVirt only maps the Tx\:Data region if needed for cache
    management (shown as dashed). We only show a single
    client of the virtualiser.}
  \label{f:regions}
\end{figure}

Note that the driver only handles \emph{pointers} to DMA buffers, it has no
need to access the actual data.  As shown in in the right half of \autoref{f:regions},
we make this explicit by separating the
\emph{data region}, which contains the buffers to be filled/emptied by
the device, from the \emph{device metadata region}, which contains the
HW-defined ring buffers pointing to the data buffers.
\emph{The data region is not mapped into the driver's address space}, in line with
the principle of least privilege. The device control region is mapped
to the driver \emph{uncached}.

In addition there is the
\emph{device control region}, which maps the device registers for
memory-mapped I/O. Data and device
metadata regions are memory regions accessed by the device via DMA.

\subsubsection{Virtualisers}

Drivers abstract the device hardware; in line with separation of
concerns they do not deal with sharing the device between multiple
clients, and the address translations required for this. This is the
responsibility of a separate \emph{virtualiser} (Virt) component which:
\begin{itemize}
\item shares a physical device between multiple clients;
\item translates references to DMA buffers from client addresses to
  device addresses (physical addresses or IOMMU-translated I/O-space
  addresses);
\item performs cache management (flushing/invalidating) where needed
  (not required on the x86 architecture, which keeps caches coherent with DMA).
\end{itemize}

For the Ethernet device class, we have independent virtualisers for
the Tx and Rx paths (TxVirt and RxVirt).

The Virt replicates its driver interface at the client side, meaning
each of the Virt's client-side interfaces looks exactly like its device
interface. Specifically it has a metadata region (Tx~V~MD, Rx~V~MD)
that structurally replicates the respective device metadata
region. The key difference is that while the device MD regions use I/O
addresses for referring to data buffers, the Virt MD regions refers to
data buffers by \emph{offsets} from the beginning of the respective
data region, thus making the Virt's address-translation task
independent of client virtual address-space layout.

The RxVirt needs to inspect the headers of incoming packets so
it has the Rx Data region mapped (R/O). The TxVirt does not need
to access the data region directly, but may need to have it mapped
to perform cache
management.  For example, the Arm architecture performs cache
operations on virtual address ranges, so the TxVirt needs the data
region to be mapped into its address space.  Arm does not have
cache-coherent DMA, so memory that will be transferred to a device
using DMA has to be cache-cleaned before DMA occurs.
Likewise, the RxVirt on Arm needs to invalidate caches after DMA into
its buffers.
These mappings are indicated in the
left half of \autoref{f:regions}.

The TxVirt must implement a \emph{traffic shaping} policy if its clients
generate load that exceeds the NIC's Tx capacity. In line with our
principle of use-case specific policy, the Tx policy is as simple as
the specific use-case allows, eg.\ round-robin,
priority-based or bandwidth limiting. %

The RxVirt at most requires a simple policy: what to do when data
arrives for a client whose RxA queue is full. Possible choices are to
block, or (more likely) discard the packet and return the buffer to the
driver's RxF queue. We generally avoid this case by ensuring that all
client queues are large enough to hold all available Rx buffers,
starving the device of buffers if the clients fail to process input
fast enough -- this leads to the NIC dropping packets under overload without wasting
CPU cycles, and leaves the RxVirt policy-free.

\subsubsection{Data regions and copiers}\label{s:copy}

The Tx data region is seen as a single region by the driver. However,
each client has its own sub-region, which is mapped into the client
address space (and the Virt's where required).

While each client data region is contiguous in physical memory, there
is no need for contiguity of the overall data region. Obviously, the
whole region must be mapped to the device by the IOMMU.

The same approach does not work for Rx data, as the device will
deposit input in any free buffer, and only the Virt determines the
target address space. There are three possible approaches to making Rx
data available to the correct client:
\begin{enumerate}
\item have only a single, global Rx data region, and the Virt maps
  each buffer to the client when inserting it into the client's RxA
  queue, and unmapping it when retrieving a buffer from the client's
  RxF queue. This needs additional privilege in the Virt, but the Virt
  must be trusted anyway;
\item\label{i:none} have only a single, global Rx data region, which is mapped R/O
  in all client address spaces. This implies that clients can read
  other client's input data;
\item\label{i:copy} have an explicit \emph{copier} (Copy) component between the
  RxVirt and each client, which copies the data from the global data
  region into a per-client data region.
\end{enumerate}

The actual choice comes down to performance (incl.\ a trade-off
between the cost of copying and the cost of mapping operations) and
the system's security policy.
For example, option (\ref{i:none})
is suitable if there is no concern about one client seeing another's
input data (e.g.\ when all network traffic is encrypted) and clients
can be trusted to return buffers.
The Copy component can be inserted
transparently: the difference between case (\ref{i:none}) and (\ref{i:copy}) above does not
affect the implementation of either the Virt or the client/copier.

\subsubsection{Broadcast}

Broadcast packets such as for the Address Resolution Protocol (ARP)
need special handling.  We offer two schemes.

The first approach uses a separate ARP client, whose only job is to
respond to those requests. This requires that incoming traffic is routed to
clients based on MAC address, and that the ARP client has (R/O)
access to the MAC-address allocation table.  Other broadcast packets
are dropped.

The alternative approach handles a broadcast packet in the RxVirt by
enqueuing a copy of the packet in each client's queue, and
reference counting the driver's buffer containing the packet. We
decrement the reference count when the buffer is returned to the
client's RxF queue; and return the buffer to the driver's RxF queue
once all clients have finished (i.e.\ the reference count has dropped
to zero).
Each client is then responsible for handling any
broadcast traffic it receives.

\subsubsection{Other device classes}\label{s:dev-oth}

Some other device classes look similar to Ethernet at a high level,
and result in a similar design. This includes most serial devices
(serial ports, SPI, I2C) with differences in the details for the
protocol. Some have no separation between data and metadata (the
queues directly contain the data).

Others, especially storage devices, do not have the clear separation
of Tx and Rx traffic of Ethernet, and instead
react to explicit requests. This results in a slightly
different design:
\begin{itemize}
\item there is a single driver metadata region;
\item there are only two queues, the \emph{request} (Rq) and the
  \emph{response} (Rs) queue;
\item there is a single Virt, which presents an Rq and Rs queue to
  each client in a per-client metadata region
\item there is one data region per client.
\end{itemize}

For storage, in addition to read and write requests, the Rq may also contain
\emph{barrier} requests, across which the device is not allowed to
reorder other requests. Other protocol details support batching of requests.
The storage Virt statically partitions the devices between its clients.

The storage driver exports an \emph{information page} of device
properties, which is appropriately virtualised by the Virt.

\subsection{OS services}

For best performance, \lions presents a \emph{native API} that is
asynchronous and modelled largely on the device interfaces. For
developer convenience and to ease porting of legacy applications,
we use a coroutine library that implements a POSIX-like \emph{blocking API}
that is layered over the native one.

As network traffic is explicitly (de)multiplexed by the virtualisers,
there is no need for a global IP stack, it becomes a library linked
directly into the client. \emph{This takes the complex (and probably buggy)
protocol stack out of the system's TCB.}

We use the same approach for storage, by providing a per-client
file-system library that directly operates on the virtual storage
device provided by the Virt (with an optional copier in
between). Alternatively, a single file system could be used
for all clients, which then is the sole client of the storage Virt --
this shared file system would have to be trusted. We currently see no
need for this in our target domain.

Sharing across per-client file systems is enabled by an explicit
multiplexer component that connects to multiple clients. In our space,
this is generally used for read-only storage.

\section{\lions Implementation}\label{s:impl}

\subsection{The starting point: seL4 Microkit}

We base the design of \lions on the seL4 Microkit \citep{microkit:url%
} (formerly ``Core Platform'').  The Microkit simplifies seL4
usage by imposing a static system architecture and an
event-driven programming model.  It presents an abstraction of the seL4
API that is partially verified using SMT solvers \citep{Paturel_SH_23}.

The Microkit provides a process abstraction called
\emph{protection domain} (PD).  PDs are single-threaded and combine the
seL4 abstractions of virtual address space, capability space, thread
and scheduling context. Multi-threaded processes can be implemented through multiple PDs that
share an address space.  While useful for applications, we do
not use this for \lions itself -- all \lions components are strictly
sequential.

PDs communicate via shared memory and semaphores (seL4
notifications). Server-type PDs can be invoked synchronously via
\emph{protected procedure calls} (PPCs), which map onto seL4 synchronous
IPC -- such a server executes on the caller's core.

PDs are structured as event handlers. Signalling a PD's semaphore will
cause it (eventually) to execute the \code{notified} function, identifying the sender PD. A
server has another handler function, \code{protected}, to receive
PPCs. Each PD also has an \code{init} handler for initialisation.

The system architecture of PDs and their
communication channels (semaphores and shared memory regions)
is defined in a \emph{system description file} (SDF).  It
specifies the ELF files to be loaded into each PD and a PD's
meta-data, including scheduling parameters, access rights to memory
regions and caching attributes, and access to interrupts (which appear
as semaphores).  A PD can \emph{monitor a virtual machine}, in
which case it acts as a private virtual-machine monitor
(VMM) which handles virtualisation events from that VM.

Microkit tooling generates from the SDF the seL4 system calls that set
up the PDs, channels and memory regions and invokes each PD's
\code{init} function.  The tooling hides the complexities of seL4's
capability system from the developer.

\subsection{Queues and state}

The design using explicit virtualisers (for separation of concerns)
enables another important simplification: All shared-memory
communication is single producer,
single consumer (SPSC), enabling the use of simple, lock-free queue
implementations. Specifically, the TxF and RxA queues hold data that
is provided by the driver (original producer) and consumed by the
client (ultimate consumer), packets flow from right to left in
\autoref{f:regions}; for the TxA and RxF queues the flow is in the
opposite direction.

The queues are also inherently bounded, leading to a
simple, array-based implementation, where references to particular
queue entries are array indices. We also require all queues to
be a power of two in size, further simplifying implementation and
sanitation.

An important property of this design is that \emph{all policy-independent
state is held in shared memory}. This makes it easy to restart a failed
component without affecting the rest of the system (other than by a
transient latency glitch). This even enables switching
policies on the fly, by reloading the code of a
component. We demonstrate this in \autoref{s:swap}.

\subsection{Location transparency}\label{s:core-mgmt}

In standard producer-consumer fashion, the lock-free SPSC queues are
synchronised by semaphores (signalling a Microkit channel). A
producer component signals the consumer if new buffers have
been enqueued in a previously empty queue,
and the consumer has set a flag that requests
signalling.
Similarly, producers can request signalling on a queue becoming
non-full.

This approach is completely \emph{location transparent}: A particular
component is not aware whether the component with which it shares a
queue is running on the same or a different core.  This location
transparency of components makes up for the strictly sequential nature
of \lions components: Instead of requiring error-prone,
multi-threaded implementation of components to make use of multicore
hardware, \lions utilises multicore processors by distributing
components across cores.

The result is that \emph{concurrency is tamed}: almost all code is
freed from concurrency control. The only requirements are correct use
of semaphores and flags, and the correct implementation of the
enqueue/dequeue library functions (which are straightforward due
to the SPSC nature of the queues).

Location transparency will also simplify core management (the
implementation of which is in progress): If a core
needs to be off-lined, components running on it can be transparently
migrated to other cores, without affecting the system's operation
(other than some temporary latency increases).

\subsection{Legacy driver reuse}

The \lions design vastly simplifies drivers compared to other OSes
(see \autoref{s:complexity}); implementing drivers from scratch
is usually easy and
will result in the best performance.

However, it is unrealistic to expect adopters to write all drivers
from scratch, especially since in practice few devices are performance
critical enough to justify such an effort. It is also frequently
impractical, as many devices are poorly (or un-)documented. For such
cases, \lions allows reusing a driver from Linux by
encapsulating it in a virtual machine (VM). Unlike the Dom0 driver VM
of Xen \citep{Barham_DFHHHNPW_03} or the Driver Container of HongMeng
\citep{Chen_JWLLLWHLYWYPX_24}, we follow the approach of
\citet{LeVasseur_USG_04} and support (but do not force) wrapping each driver in its own VM.

\begin{figure}[t]
  \center
  \includegraphics[width=0.95\linewidth]{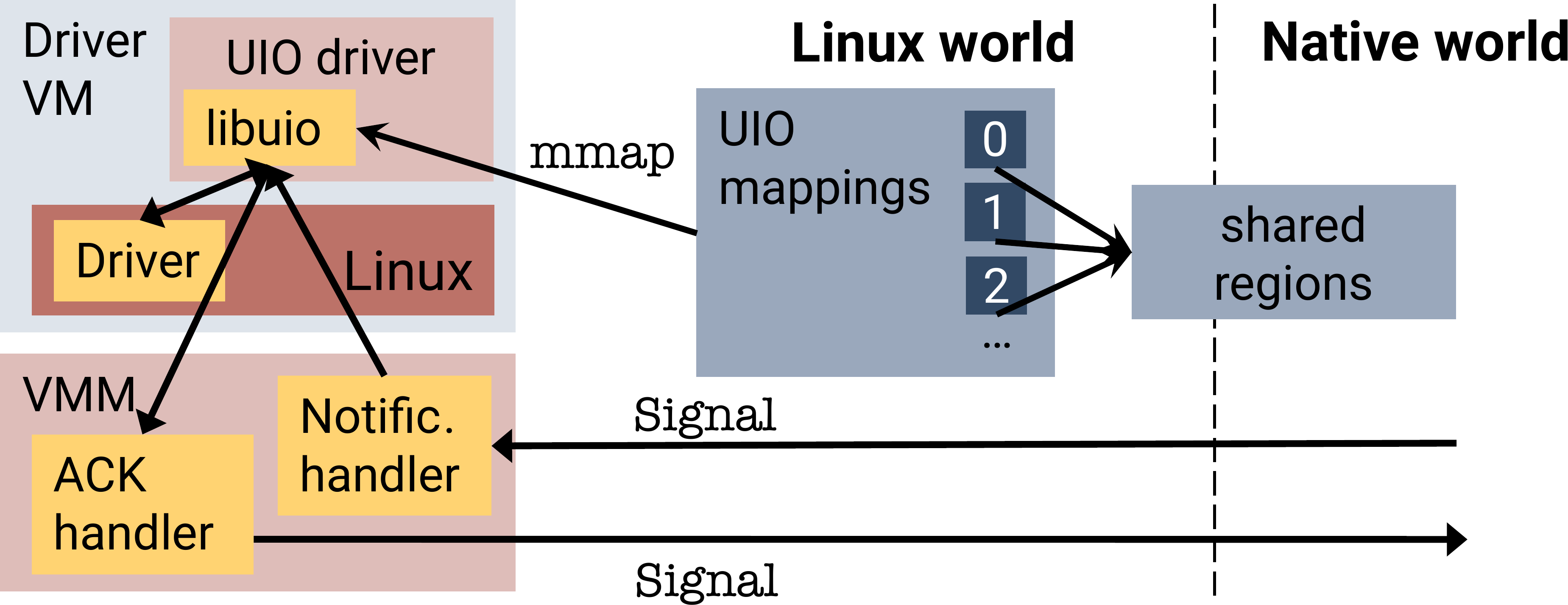}
  \caption{Driver-VM architecture}
  \label{f:drivervm}
\end{figure}

\autoref{f:drivervm} shows the architecture. The driver VM runs the
legacy driver as part of a (minimally configured) Linux guest. The
guest runs a single, statically-linked usermode program, the \emph{UIO
  driver} (which replaces \code{init}). The program uses normal Linux
system calls to interact with the device, and the Linux user I/O (UIO)
framework to interact with the \lions driver queues.

Specifically we use UIO to map guest physical memory (to access the
queues) and receive virtual interrupts. seL4's virtual machine
architecture re-directs virtualisation exceptions to a per-VM
virtual-machine monitor. We use this to inject semaphore signals
from the Virt as IRQs into the VM, to be received by the UIO driver.

We supply the driver VM's complete userspace as a CPIO archive loaded
at boot time from a RAM disk.

\subsection{Implementation status}

\subsubsection{Device drivers}

Most of our development happens on the HardKernel Odroid-C4
(Amlogic S905X3 SoC) and the Avnet MaaXBoard (i.MX8MQ SoC) platforms,
so this is where we currently have the largest set of native
drivers:
\begin{itemize}
\item Serial for all supported platforms.
\item PinMux and clock for MaaXBoard and Odroid-C4.
\item Ethernet for Odroid-C4, MaaXBoard and the i.MX8 series FEC.
\item Block: SDHC drivers for the MaaXBoard and Odroid-C4 -- the
  latter written in Rust.
\item VirtIO drivers (for running on top of QEMU) for serial, block,
  network and graphics (2D).
\item an \ISC host driver for the Odroid-C4.
\item \ISC drivers (using the \ISC host driver) for a PN532 NFC card reader and a
  DS3231 real-time-clock.
\end{itemize}
Most drivers are written in C, but the system does
not prescribe an implementation language, as demonstrated by the
Odroid-C4 Rust-implemented SDHC driver.

We have Linux driver VMs for the following device classes:
\begin{itemize}
\item GPU via exported framebuffer for Odroid-C4.
\item Ethernet for Odroid-C4.
\item Block (SDHC) for the Odroid-C4.
\item Sound using the ALSA framework for Odroid-C4.
\end{itemize}

\subsubsection{Services}\label{s:kitty}

We have full networking functionality as described above, using
lwIP \citep{Dunkels_01} as a client library: lwIP cannot break
isolation and is thus not part of the system's TCB.  Our default setup uses an RxCopy component (see \autoref{s:copy}).
Both the
native (asynchronous) as well as the blocking API are supported, the
latter is layered over the former using a coroutine library. We
have an NFS client (using an open-source NFS library) which uses the
blocking API.

We have an asynchronous filesystem API that uses either a native FAT
file system, the NFS client for network storage, or any Linux file
system hosted in a Linux VM accessed via the
standard VirtIO block interface. Again, there is a blocking API
layered on top.

\lions is mature enough to run several complete
systems in daily use. One of them is a reference design for a
point-of-sales terminal. It uses a driver VM to re-use the Linux
GPU driver, and either a native Ethernet driver or another driver VM
for re-using a Linux driver (to demonstrate multiple driver VMs).
\begin{figure}[t]
  \centering
  \includegraphics[width=0.9\linewidth]{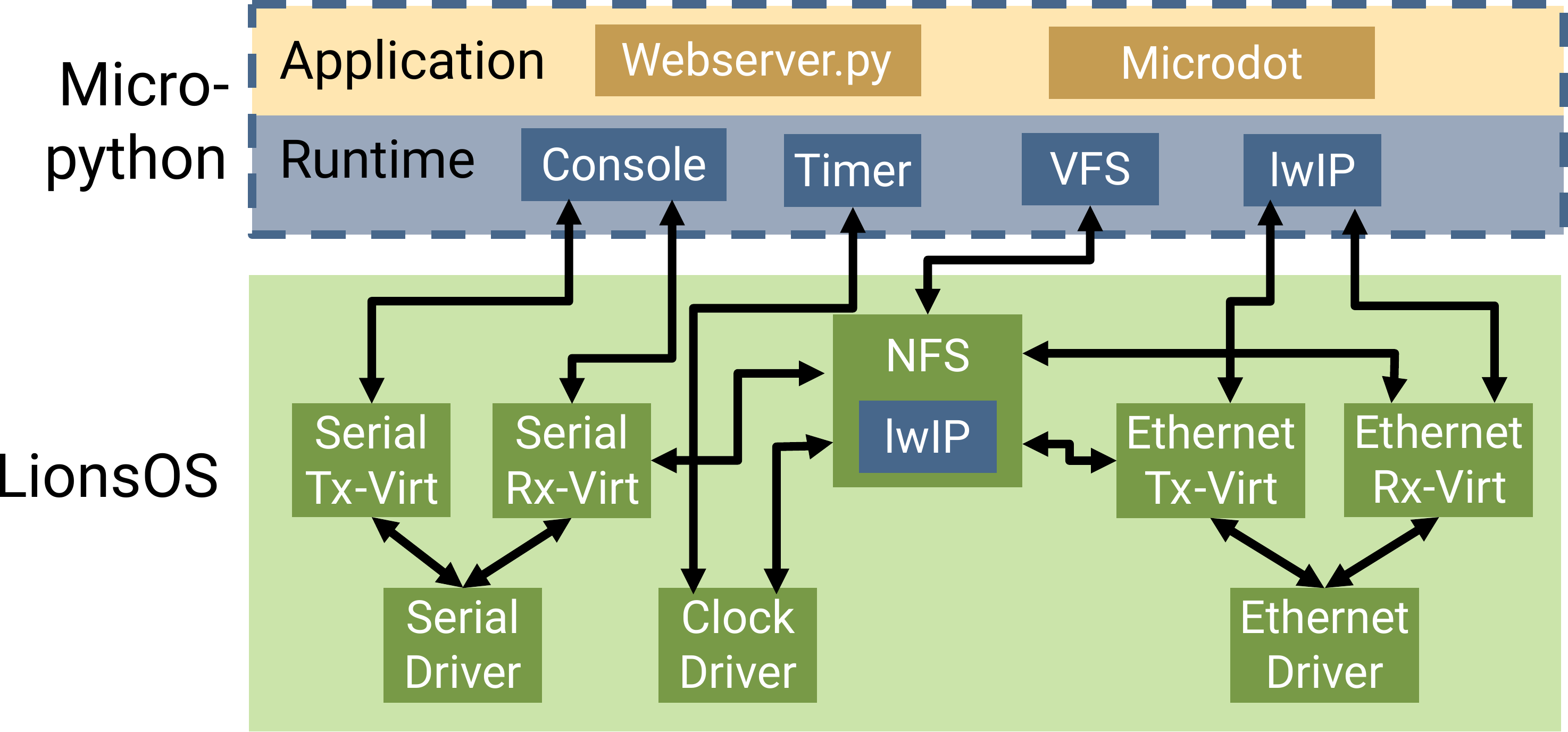}
  \caption{Architecture of the \lions-based web server.}
  \label{f:webserver}
\end{figure}

Another deployed system is a web server that hosts
the \href{https://sel4.systems}{sel4.systems}
web site.
The web server has the business logic implemented in Python,
supported by a port of MicroPython \citep{upython:url} to \lions,
\autoref{f:webserver} shows the architecture.

\subsubsection{Resource management}

Implementing dynamic resource management, such as core
off-/on-lining (cf. \autoref{s:core-mgmt}), is in progress.

\section{Evaluation}\label{s:eval}

We evaluate multiple aspects of \lions, covering development and
debugging effort, legacy driver re-use, and performance.

\subsection{Platforms}

\lions supports the Arm AArch64, Intel x86\_64 and RISC-V RV64
architectures. Development happens primarily on Arm
platforms and is then ported to the other architectures. We evaluate on
Arm and x86.

The Arm platform is an Avnet MaaXBoard with an NXP i.MX8MQ SoC,
having four Cortex A53 cores running at a maximum of 1.5\,GHz and
sharing an L2 cache. We run
all measurements at a clock rate of 1\,GHz to prevent overheating.
The board has 2\,GiB of RAM, an on-chip
1\,Gb/s NIC, and an on-chip SDHC controller.  We perform Linux
measurements on this board
with a small \citet{Buildroot:URL} system using kernel version 6.1.0.
Some measurements are made on a HardKernel Odroid-C4, which has an
AMlogic S905X3 quad-core Cortex A55 running at 1.2\,GHz, 4Gb of DDR4
RAM, and on-chip NIC and SDHC devices.

The x86 platform is an Intel Xeon\textsuperscript\textregistered{} W-1250 six-core CPU running at
3.3\,GHz, private L2 caches and a shared L3.
It has an Intel IXGBE X550 10\,Gb/s copper NIC.  We disable
hyperthreading and turbo-boost.  For Linux measurements we use a Debian
\emph{Bullseye} userspace and use the ``performance''  CPU frequency controller
to keep the CPU frequency at its maximum value.  The kernel is the Debian Linux kernel 6.6.15-2 (2024-02-04) running the standard
in-kernel IP stack.

\subsection{Complexity and development effort}\label{s:complexity}

\begin{table}[b]
  \centering
  \begin{tabular}{|l|r|r|r|r|}
    \hline
    Platform & Speed & Linux & \lions & Ratio \\
    \hline \hline
    i.MX8 & 1\,Gb/s & 4,775 & 569 & 8.4 \\
    S905X3 & 1\,Gb/s & 10,678 & 397 & 26.9 \\
    x86 & 10\,Gb/s & 3,019 & 668 & 4.5 \\
    \hline
  \end{tabular}
  \caption{SLOC of \lions NIC drivers compared with Linux.}
  \label{t:drivers}
\end{table}

\subsubsection{Code size}

Our subjective experience is that the \lions model dramatically
simplifies development of core OS components. A striking example is
the i.MX8 network driver, which was the first device
driver written to the \lions driver model of \autoref{s:des-dd}. It
was implemented by a second-year undergraduate
student, less than 18 months after she wrote her first
program. \autoref{t:drivers} compares the code sizes of Ethernet
drivers to Linux. We use \code{sloccount} \citep{Wheeler_01} for all code-size
measurements.

The size gives an indication of the complexity of the task. The
student found she was spending very little time in debugging the
driver logic, unlike normal driver development.

\newcommand{\nTCB}{\color{accent}}
\begin{table}[b]
  \centering
  \begin{tabular}{|lr|lr|}
    \hline
    \bf Component & \bf LoC & \bf Library & \bf LoC \\
    \hline
    \hline
    Serial Driver & 249 & Microkit & 303 \\
    \hline
    Serial TxVirt & 175 & Serial queue & 219 \\
    \hline
    Serial RxVirt & 126 & \ISC queue & 101 \\
    \hline
    \ISC Driver & 514 & Eth queue & 140 \\
    \hline
    \ISC Virt & 154 & \multirow{2}{20mm}{\raggedleft Filesys queue \& protocol}  & \\
    \cline{1-2}
    Timer Driver & 136 && \raisebox{1.5ex}[0mm][0mm]{268} \\
    \hline
    Eth Driver & 397 & \nTCB  Coroutines & \nTCB 848\\
    \hline
    Eth TxVirt & 122 & \nTCB lwIP & \nTCB 16,280 \\
    \hline
    Eth RxVirt & 160 & \nTCB NFS & \nTCB 45,707\\
    \hline
    Eth Copier & 79 & \nTCB VMM & \nTCB 3,098  \\
    \hline
    \hline
    \bf Total & 2,112 & \multicolumn{2}{r|}{1,031 + \nTCB 65,933} \\
    \hline
  \end{tabular}
  \caption{Code-size breakdown of the point-of-sale terminal
    demonstrator. {\nTCB Components in red font} are not part of the
    \lions TCB. The lwIP and NFS libraries ported from other
    systems, the rest is written from scratch.}
  \label{t:kitty}
\end{table}
Our experience with other \lions components is similar, components are
small and simple. \autoref{t:kitty} gives a breakdown of the
Odroid-C4-based point-of-sales terminal (see
\autoref{s:kitty}). \lions, as configured for this application,
consists of about 3.1\,kSLOC of trusted code, plus 66\,kSLOC of untrusted
library code that cannot break isolation. All the trusted code is
written from scratch.

\subsubsection{Signalling protocols}

While the \lions design clearly reduces overall complexity, it does
shift some complexity into the inter-component synchronisation
protocols. A pessimistic implementation of those protocols is trivial:
the producer notifies the consumer whenever it inserted something into
a shared queue, and the consumer signals whenever it took something
out of a queue. This clearly leads to over-signalling and should be
avoided.

A potential refinement is to signal only if there is a substantial
change to a queue: The producer adding to an empty queue or the
consumer removing from a full queue. However, race conditions can
result in subtle protocol bugs that lead to deadlocks.

Fortunately we find that model-checking is effective at eliminating
these bugs. We create models of each
our components and use SPIN \citep{Holzmann_97} to prove
deadlock freedom. This allows us to
optimise each component's signalling decision.

\subsubsection{Driver VMs}

Contrary to claims made by \citet{Chen_JWLLLWHLYWYPX_24} (which are
based on work by \citet{LeVasseur_USG_04} that preceded hardware
support for virtualisation), we find that
reusing a Linux driver in a virtual machine is an easy way to gain
access to a driver.  The main challenges are platform initialisation, and
avoiding excessive resource usage at runtime.

\paragraph{Effort}

During the development process, the VM image can include standard
Linux development and debugging tools, supporting rapid development
of the UIO driver, typically within a few days.
Once built, that UIO driver will work for \emph{any} device of the class it
has been built for.

The resulting engineering effort per device-class is small (a few weeks).  The main
disadvantage (apart from the relatively large size of an entire VM for
a single driver) is increased latency in accessing the device, so this
is unlikely to be an option for performance-critical devices.
In practice, the cost of interacting with a driver only matters for a
small number of devices (mostly networking, storage and busses).

\paragraph{Reducing resource usage}

\begin{table}[b]
  \center
  \begin{tabular}{|l|r|r|r|}\hline
    \textbf{Driver} & \textbf{Kernel} & \textbf{RAM disk} & \textbf{Runtime} \\
    \hline
    Default config  & 29 & 6.7 & 70\\ \hline
    GPU & 33 & 6.4 & 128 \\ \hline
    Audio & 3 & 2.4 & 18\\ \hline
    Block & 3 & 0.05 & 12\\ \hline
  \end{tabular}
  \caption{Sizes of driver VMs in MiB (GPU is \textbf{not} optimised).}
  \label{t:driverVM-sizes}
\end{table}

RAM usage is reduced by carefully configuring a minimal kernel and
userspace, see
\autoref{t:driverVM-sizes}. The
default configuration is what the \texttt{buildroot} tool provides for
Odroid-C4; the unoptimised GPU configuration has extra memory mapped
in for the framebuffer device communication, and is built with all
drivers compiled in (rather than modules in the file system).
The optimised audio and
block device VMs show the order-of-magnitude gains possible with a bit of effort. The
memory footprints of the optimised VMs are small given today's memory
sizes (but co-locating uncritical drivers in a single VM is always possible).

\paragraph{Platform initialisation}

Most devices need various platform registers to be configured,
to set up clocks for the device, and connections to the outside world
(pins).
We use native PinMux and clock drivers to set up the
platform to enable all devices in the device tree.  Linux also wants to
control these, so the driver VM traps
accesses to the PinMux and clock registers, discards writes, and
on read, queries the native drivers, and passes the actual values in
the registers to Linux drivers.
For debugging,  the native drivers support printing the values passed to
them.

Some devices require changing PinMux or clock settings at run
time. For example Odroid-C4 uses a pin either as a clock output, or as a
GPIO, while upgrading an SDHC card from its initial low-speed to a
high-speed state. Supporting such dynamic setting is on-going.

\subsection{Performance}

The fine-grained modularity of the \lions design implies a higher
context-switching rate than for monolithic designs, making IPC
costs a potential performance bottleneck ~\citep{Mi_LYWC_19,
  Chen_JWLLLWHLYWYPX_24}.

We focus on networking, which stresses the design the most out of
all the realistic use cases we can think of: With standard 1.5\,KiB
Ethernet frames, a 1\,Gb/s NIC (typical of embedded platforms) can
handle 81,000 packets per second for transmit and the same number for
receive, resulting in 100,000s context switches per second;
a 10\,Gb/s NIC will impose ten times that load. This is at least
as challenging as the 60,000 syscalls/s quoted by
\citet{Chen_JWLLLWHLYWYPX_24} for smartphones.

We also look at storage, which can be performance-critical too
\citep{Chen_JWLLLWHLYWYPX_24}.

\subsubsection{Networking}

\begin{figure*}[t]
  \centering
  \subcaptionbox{Unicore CPU utilisation.\label{f:UDP_echo_arm_CPU}}
    {\includegraphics[width=0.25\linewidth]{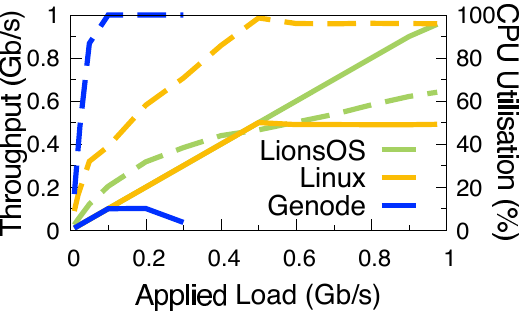}}\hspace{\fill}%
  \subcaptionbox{Unicore round-trip times.\label{f:UDP_echo_arm_rtt}}
    {\includegraphics[width=0.25\linewidth]{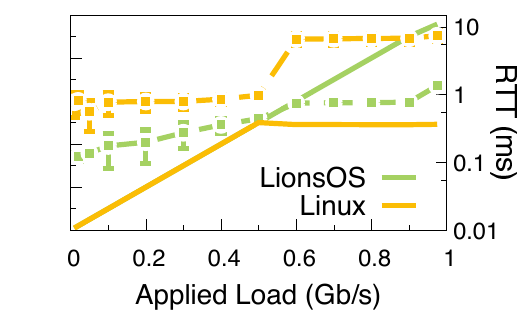}}\hspace{\fill}%
  \subcaptionbox{Multicore CPU utilisation.\label{f:UDP_echo_arm_multicore}}
    {\includegraphics[width=0.25\linewidth]{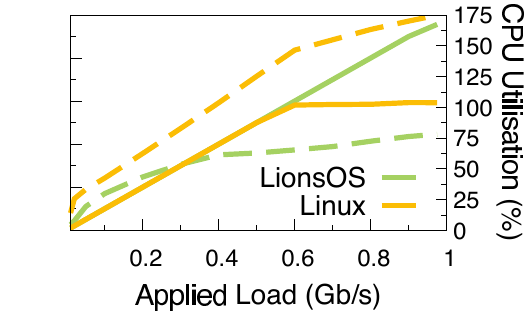}}\hspace{\fill}%
  \subcaptionbox{Multicore CEOS.\label{f:UDP_echo_ceos}}%
    {\includegraphics[width=0.25\linewidth]{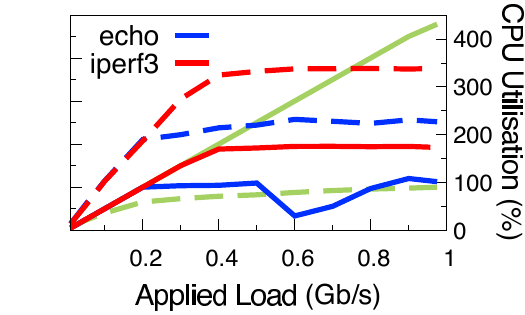}}%
  \vspace*{-2ex}
  \caption{Performance comparisons on i.MX8MQ UDP echo benchmark. Here
    and in other figures, the left axis and solid lines are throughput
    except where stated otherwise, dashed lines as per graph label (right axis).}
  \label{f:UDP_echo_arm}
\end{figure*}

To measure networking performance, we configure the evaluated system
with an \emph{echo} client, which receives packets
from the network and immediately sends them back.

We use an external load-generator machine, connected to the evaluation
platform via a switch. The load generator runs \texttt{ipbench}
\citep{Wienand_Macpherson_04} to apply a varying load of UDP traffic and measures the
achieved throughput, i.e.\ the bandwidth received back from the
evaluation system, as well as round-trip time
(RTT). We measure the CPU utilisation of the evaluated system by
running on each core a low-priority idle thread that counts free
cycles.

\begin{figure*}[t]
  \subcaptionbox{Unicore CPU utilisation.\label{f:UDP_echo_x86_CPU}}
    {\includegraphics[width=0.25\linewidth]{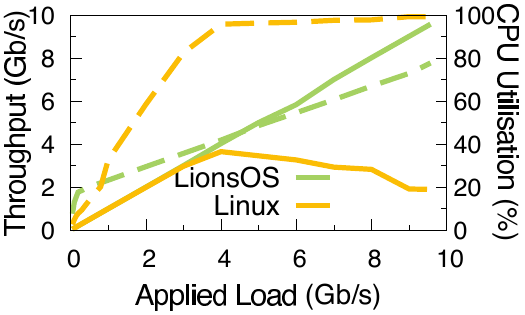}}\hspace{\fill}%
  \subcaptionbox{Unicore round-trip times.\label{f:UDP_echo_x86_rtt_xput}}
    {\includegraphics[width=0.25\linewidth]{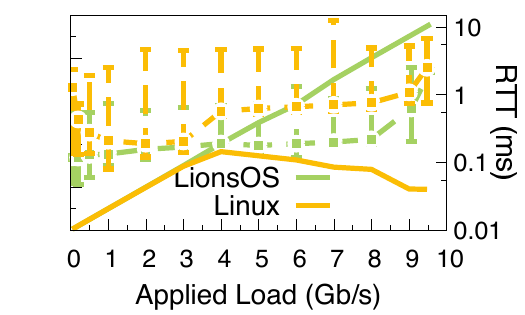}}\hspace{\fill}%
  \subcaptionbox{Multicore CPU utilisation.\label{f:UDP_echo_x86_multicore}}
    {\includegraphics[width=0.25\linewidth]{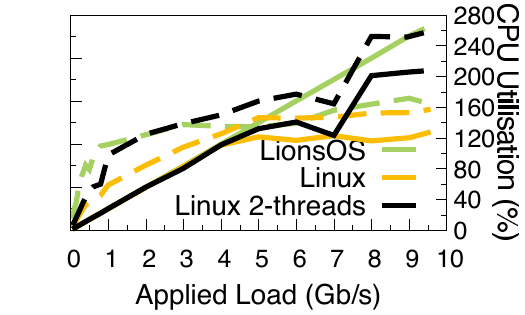}}\hspace{\fill}%
  \subcaptionbox{Simulated IPC costs.\label{f:x86_unicore_simul_cpu}}
    {\includegraphics[width=0.25\linewidth]{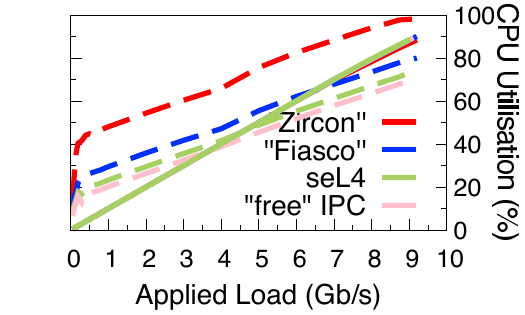}}%
  \vspace*{-2ex}
  \caption{\lions vs Linux performance on x86\_64 UDP echo
    benchmark. Dashed lines are CPU use except in graph b (RTT).}
  \label{f:UDP_echo_x86}%
\end{figure*}

\paragraph{Linux on Arm}\label{s:nw_arm}

\autoref{f:UDP_echo_arm_CPU} shows the result on the Arm platform, where
everything is pinned to a single core (ignore the blue lines for now). The \lions system has no
problem keeping up with the load, up to the capacity of the 1\,Gb/s NIC,
while Linux plateaus at about 500\,Mb/s throughput. Looking at
the CPU load we see the reason: Linux maxes out the core at an applied
load of 500\,Mb/s, while \lions achieves the full Gb/s throughput with
about one third of CPU capacity left spare. In other words, \lions uses less
than half the CPU of Linux for handling the same load.

\autoref{f:UDP_echo_arm_rtt} shows that this performance does not come
at the expense of latency -- \lions RTTs are between three and ten
times less than Linux's. In particular, we find that (unlike Linux) we
do not need to configure a Rx IRQ hold-off value to handle the
load. We do benefit from natural batching, where further packets
arrive while the driver is processing earlier ones. When it finishes, the
driver checks for new pending interrupts before going to sleep.
It will never busy-poll (Linux switches to polling under high
load).

\autoref{f:UDP_echo_arm_multicore} shows multicore results.
Here we distribute the \lions system over four cores: two cores each
run a client and its RxCopy  (only
one client is active in most benchmarks), one runs
the NIC driver and TxVirt, and the forth core runs the
RxVirt as well as other system components (timer and serial
drivers). This forces control for each packet to cross cores four
times per round-trip.  Linux is allowed to use cores as it chooses.

We see that the \lions CPU load increases from 64\% to 78\%, a
result of cross-core IPC costs being significantly higher than intra-core.
Linux still fails to handle the full load (but only uses
1.75 cores at maximum load). The performance gap between the
systems is even higher than on unicore.

\paragraph{Other OSes}\label{s:other_os}

We also run this benchmark on Genode 24.11 \citep{Feske:genode}, the
seemingly only other highly modular protected-mode OS. Only the "hw" base
configuration runs on our
platform. It is the configuration with least overhead, as it
runs bare metal without a microkernel underneath, so
some system services run in privileged mode with no IPC overhead, as
in a monolithic kernel. Key networking components, such as the IP stack,
resource multiplexers and the NIC driver, still run in userspace.

We configure
Genode with three components on unicore: the echo server (using lwIP), the uplink
component (similar to our Virts), and the NIC driver, which is transplanted from Linux
6.6.47. We take all  directly from the Genode source tree,
compose into a single system with a custom run script based on
the \code{mnt\_pocket\_stmmac\_nic} example. lwIP could be easily
swapped out for the other provided network stack, lxIP, which is ported
from Linux, but we found that this performed significantly worse in our
experiments.

The Genode throughput is the blue line in \autoref{f:UDP_echo_arm_CPU}.
It reaches a maximum of about 100\,Mb/s throughput, where it maxes out the CPU, and then
collapses at 300\,Mb/s applied load. Given this disappointing result,
we also run netperf benchmarks on the netperf\_lwip example provided
by Genode.  The TCP\_STREAM and TCP\_MAERTS benchmarks, which measure
unidirectional TCP performance, achieve around 250Mb/s and 230Mb/s
respectively, still far below throughputs achieved by either Linux or \lions.%
\footnote{We asked for help on the Genode mailing list and
  communicated directly with Genode staff, and followed their
  suggestions, which did not improve performance above what we
  report here.}

We also compare to a commercial microkernel-based operating system,
code-named CEOS.%
\footnote{The licensing conditions do not allow us to identify the system.}
CEOS does not support our standard platform (the Maaxboard.),
so we evaluate it  and \lions on the Compulab IOTGate (i.MX 8M Plus SoC), with 4 Cortex-A53
cores running at 1.2\,GHz and 6\,GiB of RAM. The board has two 1\,Gb/s NICs; in
our evaluation, we use the same FEC NIC that is used by the Maaxboard. The CEOS
network driver is provided in binary format and does not use certain hardware
features, such as Tx IRQ coalescing and HW checksum offloading, so we disable
these features in our \lions driver for fair comparison. We run the benchmark
on SMP configurations of both CEOS and \lions, as the CEOS networking architecture is optimized for
multicore and performed poorly in single-core experiments.

CEOS provides a POSIX socket API, which is the suggested method to interact
with the networking architecture, so we port our Linux echo server benchmark to it.
This achieves a maximum throughput of around 240\,Mbps, as shown by the blue line in
\autoref{f:UDP_echo_ceos}. As the echo server benchmark does not fully utilize the
CPU of the system, we also run \code{iperf3}, shown by the red
line. It is
configured as a bidirectional UDP benchmark, with CEOS acting as the server.
This achieves better throughput as CPU utilization continues to scale linearly,
maxing out at around 390\,Mbps, but performance is still far below
that of \lions (green lines).

\begin{figure*}[t]
  \centering
  \subcaptionbox{Optional components.\label{f:arm_lionsos_error_bar}}
    {\includegraphics[width=0.22\linewidth]{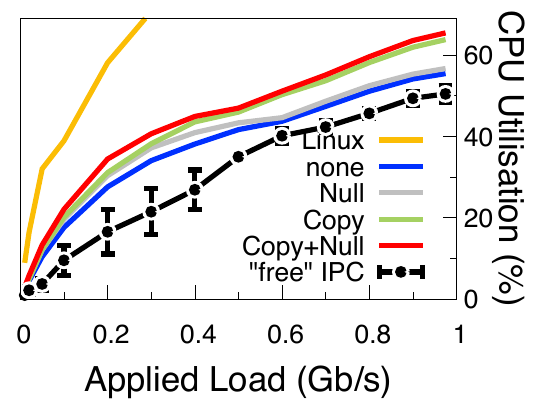}}\hspace{\fill}%
  \subcaptionbox{Multiple clients, 1 loaded.\label{f:arm_virt_scale_single_client}}
    {\includegraphics[width=0.22\linewidth]{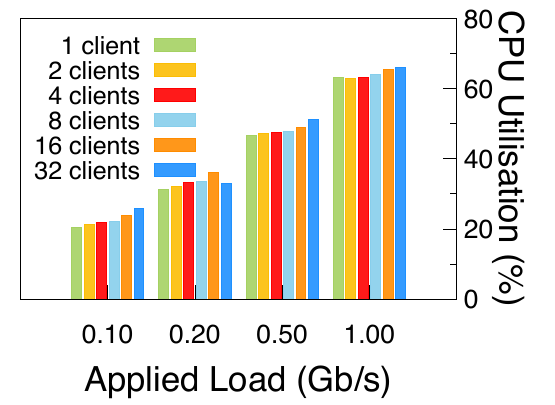}}\hspace{\fill}%
  \subcaptionbox{Multiple clients, all loaded.\label{f:arm_virt_scale}}
    {\includegraphics[width=0.22\linewidth]{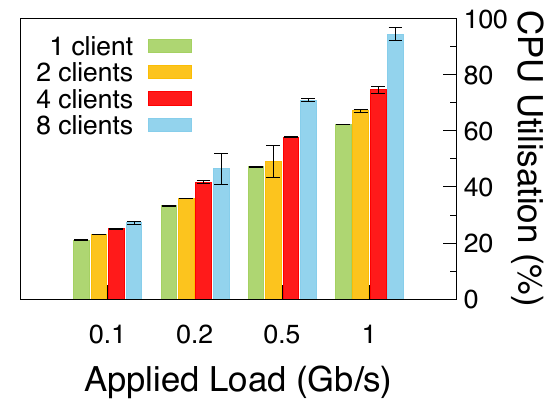}}\hspace{\fill}%
  \subcaptionbox{Virtualiser
  swap experiment.\label{f:virt_swapping}}%
    {\includegraphics[width=0.34\linewidth]{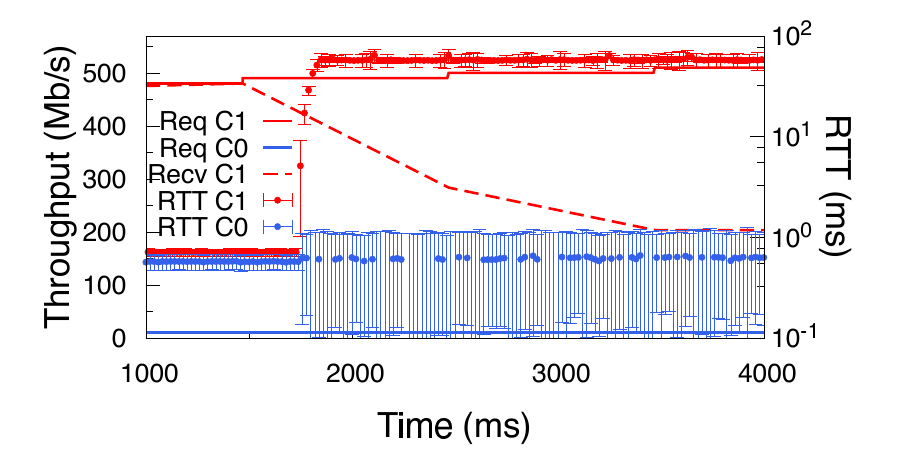}}%
  \vspace*{-2ex}
  \caption{Component performance investigations.}
  \label{f:IPC}
\end{figure*}

\paragraph{x86}\label{s:nw_x86}

As Linux might be more performance-tuned on x86 than on Arm,
we repeat these measurements on the x86 platform with a 10\,Gb/s NIC.
\autoref{f:UDP_echo_x86_CPU} shows unicore results.

\lions again handles the load with about 20\% CPU to spare, while
Linux maxes out the CPU at 4\,Gb/s load and above that its achieved
throughput deteriorates.
\lions RTTs are again much less than Linux, as
\autoref{f:UDP_echo_x86_rtt_xput} shows.

The multicore configuration, \autoref{f:UDP_echo_x86_multicore}, shows
that \lions CPU usage is almost doubled compared to single-core,
apparently a result of cross-core communication being much more
expensive on the x86 platform (with per-core L2 caches) than the Arm
platform (shared L2), but the system has no problem handling the
load. Linux, in turn, does not achieve better throughput than on
unicore (but avoids the performance collapse), despite having plenty
of CPU head space. We run Linux with a second configuration, where we
split the echo server into two components, with separate MAC
addresses, hence using separate NIC packet rings (black lines,
labelled ``2 threads'' in the figure). This almost doubles the
throughput achieved by Linux, but still leaves it behind \lions.

\paragraph{IPC cost}\label{s:context_switch}

We conduct experiments to evaluate the impact of
context-switching/IPC costs. For we first use IPC costs of the seL4, Fiasco.OC (now
called L4Re) and Google's Zircon microkernels as measured by
\citet{Mi_LYWC_19}. They found intra-core fastpath round-trip IPC latencies for
the three kernels to be 986, 2717 and 8157 cycles respectively. So we
simulate Fiasco and Zircon kernels by adding \((2717-986)/2=865\)  ("Fiasco'') and
\((8157-986)/2=3585\) (``Zircon'') respectively to each seL4 system
call (by executing a tight loop while monitoring the cycle
counter). We do not claim that this is in any way precise, but it
should give a rough estimate of how the same setup will perform on
those kernels. We also estimate performance with ``free''
IPC by \emph{subtracting} \(986/2=493\) cycles from every system call.

\autoref{f:x86_unicore_simul_cpu} shows
that while higher IPC costs lead to higher CPU load,
all three kernels manage to handle the applied load (barely in the
case of ``Zircon''). Making IPC free has minimal impact on
\lions performance.

\autoref{f:arm_lionsos_error_bar} shows a second experiment, where we compare our standard
unicore configuration (with the RxCopy component, ``Copy'' in the
graph) to a configuration without Copy (``none'') or  with a ``Null''
component that simply forwards packets without any processing. The
difference between Null and Copy is the cost of copying the packet,
while ``none'' vs.\ ``Null'' shows the context-switch overhead, as
does comparing ``Copy'' with ``Copy+Null''. As the figure shows, the
cost of an extra context switch is almost invisible.

For the \emph{``free'' IPC} case we subtract an
estimate of the kernel costs, simulating an ideal case where
module switches are totally free (in reality there would still be
function-calls). The difference is highest at low load (natural
batching reduces the context-switch rate at high loads, up to two
dozen packets are handled per invocation). The difference between this
over-idealised case and the normal configuration is still less than the
difference between the latter and Linux.

\begin{figure*}[t]
  \centering
  \subcaptionbox{Sequential read.\label{f:seq_block_read}}
    {\includegraphics[width=0.2\linewidth]{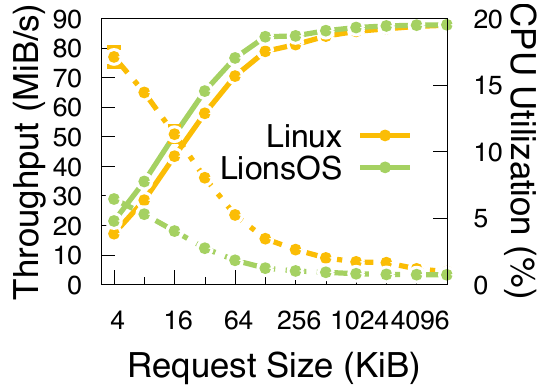}}\hspace{\fill}%
  \subcaptionbox{Random read.\label{f:rand_block_read}}
    {\includegraphics[width=0.2\linewidth]{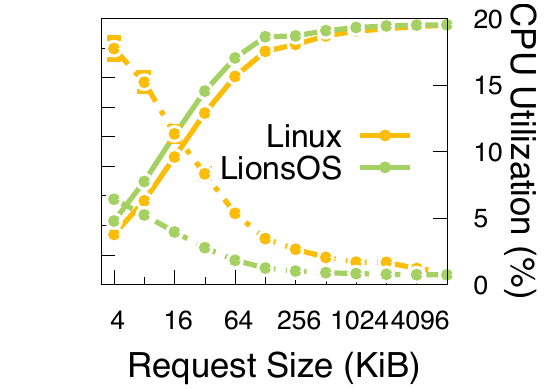}}\hspace{\fill}%
  \subcaptionbox{Sequential write.\label{f:seq_block_write}}
    {\includegraphics[width=0.2\linewidth]{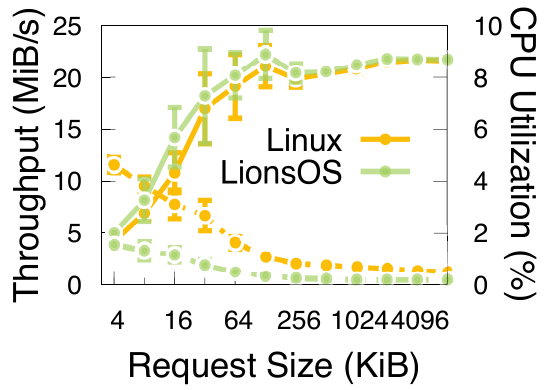}}\hspace{\fill}%
  \subcaptionbox{Random write.\label{f:rand_block_write}}%
    {\includegraphics[width=0.2\linewidth]{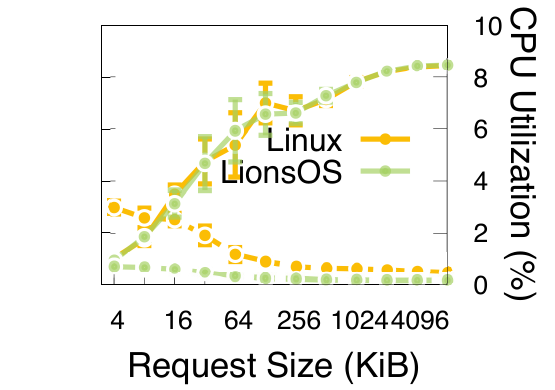}}\hspace{\fill}%
  \subcaptionbox{Reading multiple files.\label{f:fs_latency}}%
    {\includegraphics[width=0.2\linewidth]{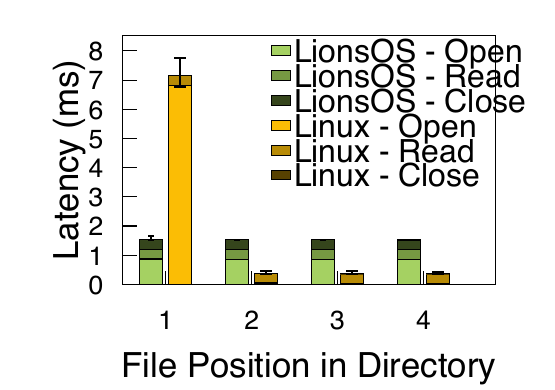}}%
  \vspace*{-2ex}
  \caption{I/O operation bandwidth on \lions and Linux. Solid lines are throughput, dashed lines are CPU utilisation.}
  \label{f:block_read_write}
\end{figure*}

\paragraph{Virtualiser}

To evaluate the scalability of sharing the NIC between multiple
clients, we vary the number of clients served by the
Virts from one to 32. \autoref{f:arm_virt_scale_single_client} shows CPU usage
as a function of applied load, when only one client has traffic
directed to it. There is only a slight dependence of CPU use on the
number of clients, resulting from the TxVirt having to poll all queues.

\autoref{f:arm_virt_scale} shows the same scenario, except that the
load is now evenly distributed between the clients. The performance
impact of client number is higher than in the previous case,
presumably a result of the higher cache footprint of the multiple
clients. In embedded systems, the number of network clients will be
small, so this is an encouraging result.

\paragraph{Dynamic policy swap}\label{s:swap}

We investigate swapping policies at run time. To demonstrate this we
configure a system with a TxVirt that monitors bandwidth, plus a
\emph{Swapper} that can reload the TxVirt with a different executable,
a bandwidth limiter, which is on standby. When a preset throughput threshold is exceeded,
the Virt signals the Swapper, which then loads the new program into the
Virt. %

\autoref{f:virt_swapping} shows the system running on the Arm
platform. We have two clients running in the system. Client~0
has a constant applied load of 10\,Mb/s, while
Client~1 has an increasing load. When this reaches a
threshold of 500\,Mb/s at about 1.7\,s, Client~1 is throttled to
200\,Mb/s through a virtualiser swap. The graph clearly shows the bandwidth limitation working as
intended, Client~0 only experiences an increased variance in its latency.
It takes 17\,\(\mu\)s to perform the switch, which is so short that
Client~0 does not experience dropped packets or even a temporary latency spike.

{\samepage
\subsubsection{Storage}
\paragraph{Raw I/O}
}
We evaluate storage-driver and Virt performance on the Odroid-C4 board
using a Sandisk SDXC card. To minimise the effect of the device's
flash management, we use a different area of the card for each
benchmark run, separated by \(>\)16\,MiB.
We power-cycle the card between runs, to clear any in-card
cache.

As shown in \autoref{f:block_read_write},
throughput of \lions and Linux are
similar, and limited by the card's performance.
However, for small block sizes, where software overheads matter,
\lions achieves this with a fraction of the CPU used by Linux.

\paragraph{File system}

The performance of reading a number of small files can matter in some
systems \citep{Chen_JWLLLWHLYWYPX_24}. We therefore conduct a
benchmark where we open-read-close a number of files on a FAT32 filesystem
using blocking I/O on both \lions and Linux running on the Odroid-C4
board.

The benchmark reads 4 files of size 512 bytes, all in the same
directory two levels from the root directory. \autoref{f:fs_latency} shows
the results compared to Linux. The time for reading each file is the
same in both systems. Linux takes a long time to open the
first file, but this is amortised when accessing multiple files --
showing the cost and benefits of directory entry caching. \lions presently
does no directory caching, and is slower when reading
more than five files. \lions takes far longer than Linux to close files, a
result of all resource management being done synchronously, while
Linux defers clean-up. %

\section{Discussion}

Our evaluation shows that \lions is able to achieve excellent
performance in networking, comprehensively outperforming Linux in
achieved throughput, latency and processing costs.
This performance is achieved despite each packet being handed between
several modules, each time requiring a context switch. Our analysis of
the impact of context-switch/IPC costs (\autoref{s:context_switch})
shows that these only have a minor impact on overall performance, and
the additional benefit from a hypothetical zero-cost IPC cost is small
compared to the already demonstrated performance advantage over
Linux. This contradicts claims in the literature that IPC cost
is still a bottleneck in microkernels \citep{Chen_JWLLLWHLYWYPX_24}.

Storage I/O performance shows the same picture: Raw I/O bandwidth equals Linux' (or is
slightly better) with a significantly reduced processing cost for
small block sizes (where software overheads dominate).

The main reason for the high performance is that each component is very
simple, doing just one thing.  There is no need for any interface layers
for adapting between multiple implementations of the same
functionality; most of the components are simple straight-line code
with no callouts via function pointers. This is possible thanks to the
simplicity afforded by keeping policy implementations specialised to
the use case.

The file-read benchmarks indicate that directory caching would help
where many files are being accessed in the same directory. This is an
example where the best policy depends on the use case, and a poor
policy choice may be detrimental. As such it is an example where a
use-case-specific caching policy should be implemented as part of the
directory management.

The experiments with swapping a component at run time
(\autoref{s:swap}) show that dynamically changing between
pre-configured policies can be done very quickly (17\,\(\mu\)s on our
Arm platform).

\section{Conclusions}\label{s:concl}

Summarising our experience with building and evaluating \lions:
\begin{itemize}
\item It is possible to build an OS based on strict separation of
  concerns, without sacrificing performance.
  \item Simplicity wins -- it more than compensates for the
  high context-switch rates resulting from modularity.
\item The simplicity of design and implementation makes us confident
  that the verification aim is achievable. The from-scratch
  implemented components that make up the system's TCB are all far
  simpler than other systems, in fact, a high-performance NIC driver has already been verified
  using automated proof techniques
  \citep{Zhao_LTUTSTASSSMNH_25}.
\item Contrary to other claims \citep{Chen_JWLLLWHLYWYPX_24} we find
  that IPC cost matters little in our design.
\end{itemize}

\begin{acks}
  LionsOS was made possible through the generous support of multiple
  organisations: the UAE Technology Innovation Institute (TII) under
  the \emph{Secure, High-Performance Device Virtualisation for seL4}
  project, DARPA under prime contract FA 8750-24-9-1000; and NIO USA. Work on the
  seL4 Microkit and the seL4 device driver framework, which form the
  foundation on which \lions is built, was supported by UK's National Cyber Security
  Centre (NCSC) under project NSC-1686.

  We also thank Ben Leslie from Breakaway for the original Microkit
  design that led to the \lions event-based programming model.
\end{acks}

\section*{Availability}
\lions is open source and available at
\url{https://github.com/au-ts/lionsos/}. It has extensive
documentation at \url{https://lionsos.org/}.

\balance
{\sloppy
  \footnotesize
      \bibliographystyle{plainnat}
    \bibliography{references}


\providecommand{\noopsort}[1]{}\providecommand{\url}{\error{The bib files now
  require the `url' package!}}\providecommand{\NoRemove}{}
\begin{thebibliography}{59}


\ifx \showCODEN    \undefined \def \showCODEN     #1{\unskip}     \fi
\ifx \showDOI      \undefined \def \showDOI       #1{#1}\fi
\ifx \showISBNx    \undefined \def \showISBNx     #1{\unskip}     \fi
\ifx \showISBNxiii \undefined \def \showISBNxiii  #1{\unskip}     \fi
\ifx \showISSN     \undefined \def \showISSN      #1{\unskip}     \fi
\ifx \showLCCN     \undefined \def \showLCCN      #1{\unskip}     \fi
\ifx \shownote     \undefined \def \shownote      #1{#1}          \fi
\ifx \showarticletitle \undefined \def \showarticletitle #1{#1}   \fi
\ifx \showURL      \undefined \def \showURL       {\relax}        \fi
\providecommand\bibfield[2]{#2}
\providecommand\bibinfo[2]{#2}
\providecommand\natexlab[1]{#1}
\providecommand\showeprint[2][]{arXiv:#2}

\bibitem[\protect\citeauthoryear{Barham, Dragovic, Fraser, Hand, Harris, Ho,
  Neugebauer, Pratt, and Warfield}{Barham et~al\mbox{.}}{2003}]%
        {Barham_DFHHHNPW_03}
\bibfield{author}{\bibinfo{person}{Paul Barham}, \bibinfo{person}{Boris
  Dragovic}, \bibinfo{person}{Keir Fraser}, \bibinfo{person}{Steven Hand},
  \bibinfo{person}{Tim Harris}, \bibinfo{person}{Alex Ho},
  \bibinfo{person}{Rolf Neugebauer}, \bibinfo{person}{Ian Pratt}, {and}
  \bibinfo{person}{Andrew Warfield}.} \bibinfo{year}{2003}\natexlab{}.
\newblock \showarticletitle{Xen and the Art of Virtualization}. In
  \bibinfo{booktitle}{\emph{ACM Symposium on Operating Systems Principles}}.
  \bibinfo{address}{Bolton Landing, NY, US}, \bibinfo{pages}{164--177}.
\newblock


\bibitem[\protect\citeauthoryear{Barhorst, Belote, Binns, Hoffman, Paunicka,
  Sarathy, Scoredos, Stanfill, Stuart, and Urzi}{Barhorst
  et~al\mbox{.}}{2009}]%
        {Barhorst_BBHPSSSSU_09}
\bibfield{author}{\bibinfo{person}{James Barhorst}, \bibinfo{person}{Todd
  Belote}, \bibinfo{person}{Pam Binns}, \bibinfo{person}{Jon Hoffman},
  \bibinfo{person}{James Paunicka}, \bibinfo{person}{Prakash Sarathy},
  \bibinfo{person}{John Scoredos}, \bibinfo{person}{Peter Stanfill},
  \bibinfo{person}{Douglas Stuart}, {and} \bibinfo{person}{Russell Urzi}.}
  \bibinfo{year}{2009}\natexlab{}.
\newblock \bibinfo{title}{A Research Agenda for Mixed-Criticality Systems}.
\newblock
\newblock
\urldef\tempurl%
\url{http://www.cse.wustl.edu/~cdgill/CPSWEEK09_MCAR/}
\showURL{%
\tempurl}


\bibitem[\protect\citeauthoryear{Bershad}{Bershad}{1992}]%
        {Bershad_92}
\bibfield{author}{\bibinfo{person}{Brian~N. Bershad}.}
  \bibinfo{year}{1992}\natexlab{}.
\newblock \showarticletitle{The Increasing Irrelevance of {IPC} Performance for
  Microkernel-Based Operating Systems}. In \bibinfo{booktitle}{\emph{USENIX
  Workshop on Microkernels and other Kernel Architectures}}.
  \bibinfo{address}{Seattle, WA, US}, \bibinfo{pages}{205--211}.
\newblock


\bibitem[\protect\citeauthoryear{Bershad, Savage, Pardyak, Sirer, Fiuczynski,
  Becker, Chambers, and Eggers}{Bershad et~al\mbox{.}}{1995}]%
        {Bershad_SPSFBCE_95}
\bibfield{author}{\bibinfo{person}{Brian~N. Bershad}, \bibinfo{person}{Stefan
  Savage}, \bibinfo{person}{Przemys{\l}aw Pardyak},
  \bibinfo{person}{Emin~G{\"u}n Sirer}, \bibinfo{person}{Marc~E. Fiuczynski},
  \bibinfo{person}{David Becker}, \bibinfo{person}{Craig Chambers}, {and}
  \bibinfo{person}{Susan Eggers}.} \bibinfo{year}{1995}\natexlab{}.
\newblock \showarticletitle{Extensibility, Safety and Performance in the {SPIN}
  Operating System}. In \bibinfo{booktitle}{\emph{ACM Symposium on Operating
  Systems Principles}}. \bibinfo{address}{Copper Mountain, CO, US},
  \bibinfo{pages}{267--284}.
\newblock


\bibitem[\protect\citeauthoryear{Brinch~Hansen}{Brinch~Hansen}{1970}]%
        {Brinch_Hansen_70}
\bibfield{author}{\bibinfo{person}{Per Brinch~Hansen}.}
  \bibinfo{year}{1970}\natexlab{}.
\newblock \showarticletitle{The Nucleus of a Multiprogramming Operating
  System}.
\newblock \bibinfo{journal}{\emph{Commun. ACM}}  \bibinfo{volume}{13}
  (\bibinfo{year}{1970}), \bibinfo{pages}{238--250}.
\newblock


\bibitem[\protect\citeauthoryear{Buildroot}{Buildroot}{2016}]%
        {Buildroot:URL}
Buildroot \bibinfo{year}{2016}\natexlab{}.
\newblock \bibinfo{title}{Buildroot: Making Embedded {Linux} Easy}.
\newblock
\newblock
\urldef\tempurl%
\url{https://buildroot.org}
\showURL{%
\tempurl}


\bibitem[\protect\citeauthoryear{Burns and Davis}{Burns and Davis}{2017}]%
        {Burns_Davis_17}
\bibfield{author}{\bibinfo{person}{Alan Burns} {and} \bibinfo{person}{Robert~I.
  Davis}.} \bibinfo{year}{2017}\natexlab{}.
\newblock \showarticletitle{A survey of research into mixed criticality
  systems}.
\newblock \bibinfo{journal}{\emph{Comput. Surveys}} \bibinfo{volume}{50},
  \bibinfo{number}{6} (\bibinfo{year}{2017}), \bibinfo{pages}{1--37}.
\newblock


\bibitem[\protect\citeauthoryear{Cebeci, Zou, Zhou, Candea, and
  Pit-Claudel}{Cebeci et~al\mbox{.}}{2024}]%
        {Cebeci_ZZCC_24}
\bibfield{author}{\bibinfo{person}{Can Cebeci}, \bibinfo{person}{Yonghao Zou},
  \bibinfo{person}{Diyu Zhou}, \bibinfo{person}{George Candea}, {and}
  \bibinfo{person}{Cl\'{e}ment Pit-Claudel}.} \bibinfo{year}{2024}\natexlab{}.
\newblock \showarticletitle{Practical Verification of System-Software
  Components Written in Standard {C}}. In \bibinfo{booktitle}{\emph{ACM
  Symposium on Operating Systems Principles}}. \bibinfo{publisher}{ACM}.
\newblock


\bibitem[\protect\citeauthoryear{Chen, Shanghai, Miao, Jia, Wang, Li, Liu, Liu,
  Wang, Huang, Li, Yang, Wang, Yin, Peng, , and Xu}{Chen et~al\mbox{.}}{2024}]%
        {Chen_JWLLLWHLYWYPX_24}
\bibfield{author}{\bibinfo{person}{Haibo Chen}, \bibinfo{person}{Shanghai},
  \bibinfo{person}{Xie Miao}, \bibinfo{person}{Ning Jia}, \bibinfo{person}{Nan
  Wang}, \bibinfo{person}{Yu Li}, \bibinfo{person}{Nian Liu},
  \bibinfo{person}{Yutao Liu}, \bibinfo{person}{Fei Wang},
  \bibinfo{person}{Qiang Huang}, \bibinfo{person}{Kun Li},
  \bibinfo{person}{Hongyang Yang}, \bibinfo{person}{Hui Wang},
  \bibinfo{person}{Jie Yin}, \bibinfo{person}{Yu Peng}, \bibinfo{person}{},
  {and} \bibinfo{person}{Fengwei Xu}.} \bibinfo{year}{2024}\natexlab{}.
\newblock \showarticletitle{Microkernel Goes General: {Performance} and
  Compatibility in the {HongMeng} Production Microkernel}. In
  \bibinfo{booktitle}{\emph{USENIX Symposium on Operating Systems Design and
  Implementation}}. \bibinfo{publisher}{Usenix}, \bibinfo{address}{Santa Clara,
  CA, US}, \bibinfo{pages}{465--485}.
\newblock


\bibitem[\protect\citeauthoryear{Chen, Li, Mesicek, Narayanan, and
  Burtsev}{Chen et~al\mbox{.}}{2023}]%
        {Chen_LMNB_23}
\bibfield{author}{\bibinfo{person}{Xiangdong Chen}, \bibinfo{person}{Zhaofeng
  Li}, \bibinfo{person}{Lukas Mesicek}, \bibinfo{person}{Vikram Narayanan},
  {and} \bibinfo{person}{Anton Burtsev}.} \bibinfo{year}{2023}\natexlab{}.
\newblock \showarticletitle{Atmosphere: Towards Practical Verified Kernels in
  {Rust}}. In \bibinfo{booktitle}{\emph{Workshop on Kernel Isolation, Safety
  and Verification}}. \bibinfo{publisher}{ACM}.
\newblock


\bibitem[\protect\citeauthoryear{Cofer, Gacek, Backes, Whalen, Pike, Foltzer,
  Podhradsky, Klein, Kuz, Andronick, Heiser, and Stuart}{Cofer
  et~al\mbox{.}}{2018}]%
        {Cofer_GBWPFPKKAHS_18}
\bibfield{author}{\bibinfo{person}{Darren Cofer}, \bibinfo{person}{Andrew
  Gacek}, \bibinfo{person}{John Backes}, \bibinfo{person}{Michael Whalen},
  \bibinfo{person}{Lee Pike}, \bibinfo{person}{Adam Foltzer},
  \bibinfo{person}{Michael Podhradsky}, \bibinfo{person}{Gerwin Klein},
  \bibinfo{person}{Ihor Kuz}, \bibinfo{person}{June Andronick},
  \bibinfo{person}{Gernot Heiser}, {and} \bibinfo{person}{Douglas Stuart}.}
  \bibinfo{year}{2018}\natexlab{}.
\newblock \showarticletitle{{A} Formal Approach to Constructing Secure Air
  Vehicle Software}.
\newblock \bibinfo{journal}{\emph{IEEE Computer}}  \bibinfo{volume}{51}
  (\bibinfo{date}{Nov.} \bibinfo{year}{2018}), \bibinfo{pages}{14--23}.
\newblock


\bibitem[\protect\citeauthoryear{Dennis and Van~Horn}{Dennis and
  Van~Horn}{1966}]%
        {Dennis_VanHorn_66}
\bibfield{author}{\bibinfo{person}{Jack~B. Dennis} {and}
  \bibinfo{person}{Earl~C. Van~Horn}.} \bibinfo{year}{1966}\natexlab{}.
\newblock \showarticletitle{Programming Semantics for Multiprogrammed
  Computations}.
\newblock \bibinfo{journal}{\emph{Commun. ACM}}  \bibinfo{volume}{9}
  (\bibinfo{year}{1966}), \bibinfo{pages}{143--155}.
\newblock


\bibitem[\protect\citeauthoryear{Dunkels}{Dunkels}{2001}]%
        {Dunkels_01}
\bibfield{author}{\bibinfo{person}{Adam Dunkels}.}
  \bibinfo{year}{2001}\natexlab{}.
\newblock \bibinfo{booktitle}{\emph{Minimal {TCP}/{IP} implementation with
  proxy support}}.
\newblock \bibinfo{type}{{T}echnical {R}eport} T2001-20.
  \bibinfo{institution}{SICS}. \bibinfo{pages}{81} pages.
\newblock
\newblock
\shownote{\url{http://www.sics.se/~adam/thesis.pdf}.}


\bibitem[\protect\citeauthoryear{Elkaduwe, Derrin, and Elphinstone}{Elkaduwe
  et~al\mbox{.}}{2008}]%
        {Elkaduwe_DE_08}
\bibfield{author}{\bibinfo{person}{Dhammika Elkaduwe}, \bibinfo{person}{Philip
  Derrin}, {and} \bibinfo{person}{Kevin Elphinstone}.}
  \bibinfo{year}{2008}\natexlab{}.
\newblock \showarticletitle{Kernel design for isolation and assurance of
  physical memory}. In \bibinfo{booktitle}{\emph{Workshop on Isolation and
  Integration in Embedded Systems}}. \bibinfo{publisher}{ACM},
  \bibinfo{address}{Glasgow, UK}, \bibinfo{pages}{35--40}.
\newblock


\bibitem[\protect\citeauthoryear{F\"ahndrich, Aiken, Hawblitzel, Hodson, Hunt,
  Larus, and Levi}{F\"ahndrich et~al\mbox{.}}{2006}]%
        {Fahndrich_AHHHRL_06}
\bibfield{author}{\bibinfo{person}{Manuel F\"ahndrich}, \bibinfo{person}{Mark
  Aiken}, \bibinfo{person}{Chris Hawblitzel}, \bibinfo{person}{Orion Hodson},
  \bibinfo{person}{Galen~C. Hunt}, \bibinfo{person}{James~R. Larus}, {and}
  \bibinfo{person}{Steven Levi}.} \bibinfo{year}{2006}\natexlab{}.
\newblock \showarticletitle{Language Support for Fast and Reliable
  Message-Based Communication in {Singularity} {OS}}. In
  \bibinfo{booktitle}{\emph{EuroSys Conference}}. \bibinfo{address}{Leuven,
  BE}, \bibinfo{pages}{177--190}.
\newblock


\bibitem[\protect\citeauthoryear{Fassino, Stefani, Lawall, and Muller}{Fassino
  et~al\mbox{.}}{2002}]%
        {Fassino_SLM_02}
\bibfield{author}{\bibinfo{person}{Jean-Philippe Fassino},
  \bibinfo{person}{Jean-Bernard Stefani}, \bibinfo{person}{Julia Lawall}, {and}
  \bibinfo{person}{Gilles Muller}.} \bibinfo{year}{2002}\natexlab{}.
\newblock \showarticletitle{{THINK}: A Software Framework for Component-based
  Operating System Kernels}. In \bibinfo{booktitle}{\emph{USENIX}}.
  \bibinfo{address}{Monterey, CA, USA}, \bibinfo{pages}{73--86}.
\newblock


\bibitem[\protect\citeauthoryear{Feske}{Feske}{2015}]%
        {Feske:genode}
\bibfield{author}{\bibinfo{person}{Norman Feske}.}
  \bibinfo{year}{2015}\natexlab{}.
\newblock \bibinfo{booktitle}{\emph{Genode Foundations}}.
\newblock \bibinfo{publisher}{Genode Labs}.
\newblock
\urldef\tempurl%
\url{https://genode.org/documentation/genode-foundations/}
\showURL{%
\tempurl}


\bibitem[\protect\citeauthoryear{Feske and Helmuth}{Feske and Helmuth}{2007}]%
        {Feske_Helmuth_07}
\bibfield{author}{\bibinfo{person}{Norman Feske} {and}
  \bibinfo{person}{Christian Helmuth}.} \bibinfo{year}{2007}\natexlab{}.
\newblock \bibinfo{booktitle}{\emph{Design of the {Bastei} {OS} architecture}}.
\newblock \bibinfo{type}{{T}echnical {R}eport} TUD-FI06-07-Dezember-2006.
  \bibinfo{institution}{Technische Universit{\"a}t Dresden}.
\newblock
\urldef\tempurl%
\url{https://www.researchgate.net/profile/Christian-Helmuth/publication/233765173_Design_of_the_Bastei_OS_Architecture/links/0912f50b5c8cd18a4c000000/Design-of-the-Bastei-OS-Architecture.pdf}
\showURL{%
\tempurl}


\bibitem[\protect\citeauthoryear{Fleisch, Co, and Tan}{Fleisch
  et~al\mbox{.}}{1998}]%
        {Fleisch_CT_98}
\bibfield{author}{\bibinfo{person}{Brett~D. Fleisch}, \bibinfo{person}{Mark
  Allan~A. Co}, {and} \bibinfo{person}{Chao Tan}.}
  \bibinfo{year}{1998}\natexlab{}.
\newblock \showarticletitle{Workplace Microkernel and {OS}: A Case Study}.
\newblock \bibinfo{journal}{\emph{Software: Practice and Experience}}
  \bibinfo{volume}{28} (\bibinfo{year}{1998}), \bibinfo{pages}{569--591}.
\newblock


\bibitem[\protect\citeauthoryear{Ford, Back, Benson, Lepreau, Lin, and
  Shivers}{Ford et~al\mbox{.}}{1997}]%
        {Ford_BBLLS_97}
\bibfield{author}{\bibinfo{person}{Bryan Ford}, \bibinfo{person}{Godmar Back},
  \bibinfo{person}{Greg Benson}, \bibinfo{person}{Jay Lepreau},
  \bibinfo{person}{Albert Lin}, {and} \bibinfo{person}{Olin Shivers}.}
  \bibinfo{year}{1997}\natexlab{}.
\newblock \showarticletitle{The {Flux} {OSKit}: A Substrate for Kernel and
  Language Research}. In \bibinfo{booktitle}{\emph{ACM Symposium on Operating
  Systems Principles}}. \bibinfo{address}{St Malo, France},
  \bibinfo{pages}{38--51}.
\newblock


\bibitem[\protect\citeauthoryear{Gefflaut, Jaeger, Park, Liedtke, Elphinstone,
  Uhlig, Tidswell, Deller, and Reuther}{Gefflaut et~al\mbox{.}}{2000}]%
        {Gefflaut_JPLEUTDR_00}
\bibfield{author}{\bibinfo{person}{Alain Gefflaut}, \bibinfo{person}{Trent
  Jaeger}, \bibinfo{person}{Yoonho Park}, \bibinfo{person}{Jochen Liedtke},
  \bibinfo{person}{Kevin~J. Elphinstone}, \bibinfo{person}{Volkmar Uhlig},
  \bibinfo{person}{Jonathon~E. Tidswell}, \bibinfo{person}{Luke Deller}, {and}
  \bibinfo{person}{Lars Reuther}.} \bibinfo{year}{2000}\natexlab{}.
\newblock \showarticletitle{The {Sawmill} multiserver approach}. In
  \bibinfo{booktitle}{\emph{SIGOPS European Workshop}}.
  \bibinfo{publisher}{ACM}, \bibinfo{address}{Kolding, Denmark},
  \bibinfo{pages}{109--114}.
\newblock


\bibitem[\protect\citeauthoryear{H{\"a}rtig, Hohmuth, Feske, Helmuth,
  Lackorzy\'{n}ski, Mehnert, and Peter}{H{\"a}rtig et~al\mbox{.}}{2005}]%
        {Hartig_HFHLMP_05}
\bibfield{author}{\bibinfo{person}{Hermann H{\"a}rtig},
  \bibinfo{person}{Michael Hohmuth}, \bibinfo{person}{Norman Feske},
  \bibinfo{person}{Christian Helmuth}, \bibinfo{person}{Adam Lackorzy\'{n}ski},
  \bibinfo{person}{Frank Mehnert}, {and} \bibinfo{person}{Michael Peter}.}
  \bibinfo{year}{2005}\natexlab{}.
\newblock \showarticletitle{The {Nizza} Secure-System Architecture}. In
  \bibinfo{booktitle}{\emph{International Conference on Collaborative
  Computing}}. \bibinfo{address}{San Jose, CA, US}.
\newblock


\bibitem[\protect\citeauthoryear{Herder, Bos, Gras, Homburg, and
  Tanenbaum}{Herder et~al\mbox{.}}{2006}]%
        {Herder_BGHT_06}
\bibfield{author}{\bibinfo{person}{Jorrit~N. Herder}, \bibinfo{person}{Herbert
  Bos}, \bibinfo{person}{Ben Gras}, \bibinfo{person}{Philip Homburg}, {and}
  \bibinfo{person}{Andrew~S. Tanenbaum}.} \bibinfo{year}{2006}\natexlab{}.
\newblock \showarticletitle{{MINIX} 3: A Highly Reliable, Self-Repairing
  Operating System}.
\newblock \bibinfo{journal}{\emph{ACM Operating Systems Review}}
  \bibinfo{volume}{40}, \bibinfo{number}{3} (\bibinfo{date}{July}
  \bibinfo{year}{2006}), \bibinfo{pages}{80--89}.
\newblock


\bibitem[\protect\citeauthoryear{Hildebrand}{Hildebrand}{1992}]%
        {Hildebrand_92}
\bibfield{author}{\bibinfo{person}{Dan Hildebrand}.}
  \bibinfo{year}{1992}\natexlab{}.
\newblock \showarticletitle{An Architectural Overview of {QNX}}. In
  \bibinfo{booktitle}{\emph{USENIX Workshop on Microkernels and other Kernel
  Architectures}}. \bibinfo{address}{Seattle, WA, US},
  \bibinfo{pages}{113--126}.
\newblock


\bibitem[\protect\citeauthoryear{Holzmann}{Holzmann}{1997}]%
        {Holzmann_97}
\bibfield{author}{\bibinfo{person}{Gerard~J. Holzmann}.}
  \bibinfo{year}{1997}\natexlab{}.
\newblock \showarticletitle{The Model Checker {SPIN}}.
\newblock \bibinfo{journal}{\emph{IEEE Transactions on Software Engineering}}
  \bibinfo{volume}{23} (\bibinfo{year}{1997}), \bibinfo{pages}{279--295}.
\newblock


\bibitem[\protect\citeauthoryear{Klein, Andronick, Elphinstone, Murray, Sewell,
  Kolanski, and Heiser}{Klein et~al\mbox{.}}{2014}]%
        {Klein_AEMSKH_14}
\bibfield{author}{\bibinfo{person}{Gerwin Klein}, \bibinfo{person}{June
  Andronick}, \bibinfo{person}{Kevin Elphinstone}, \bibinfo{person}{Toby
  Murray}, \bibinfo{person}{Thomas Sewell}, \bibinfo{person}{Rafal Kolanski},
  {and} \bibinfo{person}{Gernot Heiser}.} \bibinfo{year}{2014}\natexlab{}.
\newblock \showarticletitle{Comprehensive Formal Verification of an {OS}
  Microkernel}.
\newblock \bibinfo{journal}{\emph{ACM Transactions on Computer Systems}}
  \bibinfo{volume}{32}, \bibinfo{number}{1} (\bibinfo{date}{Feb.}
  \bibinfo{year}{2014}), \bibinfo{pages}{2:1--2:70}.
\newblock


\bibitem[\protect\citeauthoryear{Klein, Elphinstone, Heiser, Andronick, Cock,
  Derrin, Elkaduwe, Engelhardt, Kolanski, Norrish, Sewell, Tuch, and
  Winwood}{Klein et~al\mbox{.}}{2009}]%
        {Klein_EHACDEEKNSTW_09}
\bibfield{author}{\bibinfo{person}{Gerwin Klein}, \bibinfo{person}{Kevin
  Elphinstone}, \bibinfo{person}{Gernot Heiser}, \bibinfo{person}{June
  Andronick}, \bibinfo{person}{David Cock}, \bibinfo{person}{Philip Derrin},
  \bibinfo{person}{Dhammika Elkaduwe}, \bibinfo{person}{Kai Engelhardt},
  \bibinfo{person}{Rafal Kolanski}, \bibinfo{person}{Michael Norrish},
  \bibinfo{person}{Thomas Sewell}, \bibinfo{person}{Harvey Tuch}, {and}
  \bibinfo{person}{Simon Winwood}.} \bibinfo{year}{2009}\natexlab{}.
\newblock \showarticletitle{{seL4}: Formal Verification of an {OS} Kernel}. In
  \bibinfo{booktitle}{\emph{ACM Symposium on Operating Systems Principles}}.
  \bibinfo{publisher}{ACM}, \bibinfo{address}{Big Sky, MT, USA},
  \bibinfo{pages}{207--220}.
\newblock


\bibitem[\protect\citeauthoryear{LeVasseur, Uhlig, Stoess, and
  G\"{o}tz}{LeVasseur et~al\mbox{.}}{2004}]%
        {LeVasseur_USG_04}
\bibfield{author}{\bibinfo{person}{Joshua LeVasseur}, \bibinfo{person}{Volkmar
  Uhlig}, \bibinfo{person}{Jan Stoess}, {and} \bibinfo{person}{Stefan
  G\"{o}tz}.} \bibinfo{year}{2004}\natexlab{}.
\newblock \showarticletitle{Unmodified Device Driver Reuse and Improved System
  Dependability via Virtual Machines}. In \bibinfo{booktitle}{\emph{USENIX
  Symposium on Operating Systems Design and Implementation}}.
  \bibinfo{address}{San Francisco, CA, US}, \bibinfo{pages}{17--30}.
\newblock


\bibitem[\protect\citeauthoryear{Levis, Madden, Polastre, Szewczyk, Whitehouse,
  Woo, Gay, Hill, Welsh, Brewer, and Culler}{Levis et~al\mbox{.}}{2005}]%
        {Levis_05}
\bibfield{author}{\bibinfo{person}{Philip Levis}, \bibinfo{person}{Sam Madden},
  \bibinfo{person}{Joseph Polastre}, \bibinfo{person}{Robert Szewczyk},
  \bibinfo{person}{Kamin Whitehouse}, \bibinfo{person}{Alec Woo},
  \bibinfo{person}{David Gay}, \bibinfo{person}{Jason Hill},
  \bibinfo{person}{Matt Welsh}, \bibinfo{person}{Eric Brewer}, {and}
  \bibinfo{person}{David Culler}.} \bibinfo{year}{2005}\natexlab{}.
\newblock \showarticletitle{{TinyOS}: An Operating System for Sensor Networks}.
\newblock In \bibinfo{booktitle}{\emph{Ambient Intelligence}}.
  \bibinfo{publisher}{Springer}, \bibinfo{pages}{115--148}.
\newblock


\bibitem[\protect\citeauthoryear{Levy, Campbell, Ghena, Giffin, Pannuto, Dutta,
  and Levis}{Levy et~al\mbox{.}}{2017}]%
        {Levy_CGGPDL_17}
\bibfield{author}{\bibinfo{person}{Amit Levy}, \bibinfo{person}{Bradford
  Campbell}, \bibinfo{person}{Branden Ghena}, \bibinfo{person}{Daniel~B
  Giffin}, \bibinfo{person}{Pat Pannuto}, \bibinfo{person}{Prabal Dutta}, {and}
  \bibinfo{person}{Philip Levis}.} \bibinfo{year}{2017}\natexlab{}.
\newblock \showarticletitle{Multiprogramming a 64 {kB} Computer Safely and
  Efficiently}. In \bibinfo{booktitle}{\emph{ACM Symposium on Operating Systems
  Principles}}. ACM, \bibinfo{pages}{234--251}.
\newblock


\bibitem[\protect\citeauthoryear{Lyons, McLeod, Almatary, and Heiser}{Lyons
  et~al\mbox{.}}{2018}]%
        {Lyons_MAH_18}
\bibfield{author}{\bibinfo{person}{Anna Lyons}, \bibinfo{person}{Kent McLeod},
  \bibinfo{person}{Hesham Almatary}, {and} \bibinfo{person}{Gernot Heiser}.}
  \bibinfo{year}{2018}\natexlab{}.
\newblock \showarticletitle{Scheduling-Context Capabilities: {A} Principled,
  Light-Weight {OS} Mechanism for Managing Time}. In
  \bibinfo{booktitle}{\emph{EuroSys Conference}}. \bibinfo{publisher}{ACM},
  \bibinfo{address}{Porto, Portugal}, 14.
\newblock


\bibitem[\protect\citeauthoryear{Matichuk, Murray, Andronick, Jeffery, Klein,
  and Staples}{Matichuk et~al\mbox{.}}{2015}]%
        {Matichuk_MAJKS_15}
\bibfield{author}{\bibinfo{person}{Daniel Matichuk}, \bibinfo{person}{Toby
  Murray}, \bibinfo{person}{June Andronick}, \bibinfo{person}{Ross Jeffery},
  \bibinfo{person}{Gerwin Klein}, {and} \bibinfo{person}{Mark Staples}.}
  \bibinfo{year}{2015}\natexlab{}.
\newblock \showarticletitle{Empirical Study Towards a Leading Indicator for
  Cost of Formal Software Verification}. In
  \bibinfo{booktitle}{\emph{International Conference on Software Engineering}}.
  \bibinfo{address}{Firenze, Italy}, \bibinfo{pages}{11}.
\newblock


\bibitem[\protect\citeauthoryear{Mi, Li, Yang, Wang, and Chen}{Mi
  et~al\mbox{.}}{2019}]%
        {Mi_LYWC_19}
\bibfield{author}{\bibinfo{person}{Zeyu Mi}, \bibinfo{person}{Dingji Li},
  \bibinfo{person}{Zihan Yang}, \bibinfo{person}{Xinran Wang}, {and}
  \bibinfo{person}{Haibo Chen}.} \bibinfo{year}{2019}\natexlab{}.
\newblock \showarticletitle{{SkyBridge}: Fast and Secure Inter-Process
  Communication for Microkernels}. In \bibinfo{booktitle}{\emph{EuroSys
  Conference}}. \bibinfo{publisher}{ACM}, \bibinfo{address}{Dresden, DE}, 15.
\newblock


\bibitem[\protect\citeauthoryear{{MicroPython Developers}}{{MicroPython
  Developers}}{2014}]%
        {upython:url}
\bibfield{author}{\bibinfo{person}{{MicroPython Developers}}.}
  \bibinfo{year}{2014}\natexlab{}.
\newblock \bibinfo{title}{{MicroPython} documentation}.
\newblock
\newblock
\urldef\tempurl%
\url{https://docs.micropython.org/}
\showURL{%
\tempurl}


\bibitem[\protect\citeauthoryear{Narayanan, Huang, Detweiler, Appel, Li,
  Zellweger, and Burtsev}{Narayanan et~al\mbox{.}}{2020}]%
        {Narayanan_HDALZB_20}
\bibfield{author}{\bibinfo{person}{Vikram Narayanan}, \bibinfo{person}{Tianjiao
  Huang}, \bibinfo{person}{David Detweiler}, \bibinfo{person}{Dan Appel},
  \bibinfo{person}{Zhaofeng Li}, \bibinfo{person}{Gerd Zellweger}, {and}
  \bibinfo{person}{Anton Burtsev}.} \bibinfo{year}{2020}\natexlab{}.
\newblock \showarticletitle{{RedLeaf}: Isolation and Communication in a Safe
  Operating System}. In \bibinfo{booktitle}{\emph{USENIX Symposium on Operating
  Systems Design and Implementation}}. 19.
\newblock


\bibitem[\protect\citeauthoryear{Nelson, Bornholt, Gu, Baumann, Torlak, and
  Wang}{Nelson et~al\mbox{.}}{2019}]%
        {Nelson_BGBTW_19}
\bibfield{author}{\bibinfo{person}{Luke Nelson}, \bibinfo{person}{James
  Bornholt}, \bibinfo{person}{Ronghui Gu}, \bibinfo{person}{Andrew Baumann},
  \bibinfo{person}{Emina Torlak}, {and} \bibinfo{person}{Xi Wang}.}
  \bibinfo{year}{2019}\natexlab{}.
\newblock \showarticletitle{Scaling symbolic evaluation for automated
  verification of systems code with {Serval}}. In \bibinfo{booktitle}{\emph{ACM
  Symposium on Operating Systems Principles}}. \bibinfo{pages}{225--242}.
\newblock


\bibitem[\protect\citeauthoryear{Nelson, Sigurbjarnarson, Zhang, Johnson,
  Bornholt, Torlak, and Wang}{Nelson et~al\mbox{.}}{2017}]%
        {Nelson_SZJBTW_17}
\bibfield{author}{\bibinfo{person}{Luke Nelson}, \bibinfo{person}{Helgi
  Sigurbjarnarson}, \bibinfo{person}{Kaiyuan Zhang}, \bibinfo{person}{Dylan
  Johnson}, \bibinfo{person}{James Bornholt}, \bibinfo{person}{Emina Torlak},
  {and} \bibinfo{person}{Xi Wang}.} \bibinfo{year}{2017}\natexlab{}.
\newblock \showarticletitle{Hyperkernel: Push-Button Verification of an {OS}
  Kernel}. In \bibinfo{booktitle}{\emph{ACM Symposium on Operating Systems
  Principles}}. \bibinfo{publisher}{ACM}, \bibinfo{pages}{252--269}.
\newblock


\bibitem[\protect\citeauthoryear{Paturel, Subasinghe, and Heiser}{Paturel
  et~al\mbox{.}}{2023}]%
        {Paturel_SH_23}
\bibfield{author}{\bibinfo{person}{Mathieu Paturel}, \bibinfo{person}{Isitha
  Subasinghe}, {and} \bibinfo{person}{Gernot Heiser}.}
  \bibinfo{year}{2023}\natexlab{}.
\newblock \showarticletitle{First steps in verifying the {seL4} {Core
  Platform}}. In \bibinfo{booktitle}{\emph{Asia-Pacific Workshop on Systems
  (APSys)}}. \bibinfo{publisher}{ACM}, \bibinfo{address}{Seoul, KR}, 7.
\newblock


\bibitem[\protect\citeauthoryear{Qu}{Qu}{2024}]%
        {Qu_24:sel4s}
\bibfield{author}{\bibinfo{person}{Ning Qu}.} \bibinfo{year}{2024}\natexlab{}.
\newblock \bibinfo{title}{{seL4} in Software-defined Vehicles: Vision, Roadmap,
  and Impact at {NIO}}.
\newblock
\newblock
\urldef\tempurl%
\url{https://sel4.systems/Foundation/Summit/2024/slides/software-defined.pdf}
\showURL{%
\tempurl}
\newblock
\shownote{Keynote at the 6\textsuperscript{th} {seL4} {Summit}.}


\bibitem[\protect\citeauthoryear{Qubes}{Qubes}{2010}]%
        {Qubes:arch}
Qubes \bibinfo{year}{2010}\natexlab{}.
\newblock \bibinfo{title}{Qubes Architecture Overview}.
\newblock
\newblock
\urldef\tempurl%
\url{https://qubes-os.org/doc/architecture/}
\showURL{%
\tempurl}


\bibitem[\protect\citeauthoryear{Rashid, Julin, Orr, Sanzi, Baron, Forin,
  Golub, and Jones}{Rashid et~al\mbox{.}}{1989}]%
        {Rashid_JOSBFGJ_89}
\bibfield{author}{\bibinfo{person}{R.F. Rashid}, \bibinfo{person}{D. Julin},
  \bibinfo{person}{D. Orr}, \bibinfo{person}{R. Sanzi}, \bibinfo{person}{R.
  Baron}, \bibinfo{person}{A. Forin}, \bibinfo{person}{D. Golub}, {and}
  \bibinfo{person}{M. Jones}.} \bibinfo{year}{1989}\natexlab{}.
\newblock \showarticletitle{Mach: a System Software Kernel}.
\newblock \bibinfo{journal}{\emph{Spring COMPCON}} (\bibinfo{year}{1989}),
  \bibinfo{pages}{176--8}.
\newblock


\bibitem[\protect\citeauthoryear{Rawson}{Rawson}{1997}]%
        {Rawson_97}
\bibfield{author}{\bibinfo{person}{Freeman~L. Rawson, {III}}.}
  \bibinfo{year}{1997}\natexlab{}.
\newblock \showarticletitle{Experience with the Development of a
  Microkernel-Based, Multiserver Operating System}. In
  \bibinfo{booktitle}{\emph{Workshop on Hot Topics in Operating Systems
  (HotOS)}}. \bibinfo{address}{Cape Cod, MA, US}, \bibinfo{pages}{2--7}.
\newblock


\bibitem[\protect\citeauthoryear{Ren, Rodrigues, Chen, Vega, Stumm, and
  Yuan}{Ren et~al\mbox{.}}{2019}]%
        {Ren_RCVSY_19}
\bibfield{author}{\bibinfo{person}{Xiang~(Jenny) Ren}, \bibinfo{person}{Kirk
  Rodrigues}, \bibinfo{person}{Luyuan Chen}, \bibinfo{person}{Camilo Vega},
  \bibinfo{person}{Michael Stumm}, {and} \bibinfo{person}{Ding Yuan}.}
  \bibinfo{year}{2019}\natexlab{}.
\newblock \showarticletitle{An Analysis of Performance Evolution of {Linux’s}
  Core Operations}. In \bibinfo{booktitle}{\emph{ACM Symposium on Operating
  Systems Principles}}. \bibinfo{publisher}{ACM}, \bibinfo{address}{Huntsville,
  Ont, CA}, 16.
\newblock


\bibitem[\protect\citeauthoryear{Rozier, Abrossimov, Armand, Boule, Gien,
  Guillemont, Herrmann, Kaiser, Langlois, Leonard, and Neuhauser}{Rozier
  et~al\mbox{.}}{1988}]%
        {Rozier_AABGGHKLLN_88}
\bibfield{author}{\bibinfo{person}{Marc Rozier}, \bibinfo{person}{Vadim
  Abrossimov}, \bibinfo{person}{Fran\c{c}ois Armand}, \bibinfo{person}{I.
  Boule}, \bibinfo{person}{Michel Gien}, \bibinfo{person}{Marc Guillemont},
  \bibinfo{person}{F. Herrmann}, \bibinfo{person}{Claude Kaiser},
  \bibinfo{person}{S. Langlois}, \bibinfo{person}{P. Leonard}, {and}
  \bibinfo{person}{W. Neuhauser}.} \bibinfo{year}{1988}\natexlab{}.
\newblock \showarticletitle{{CHORUS} Distributed Operating Systems}.
\newblock \bibinfo{journal}{\emph{Computing Systems}} \bibinfo{volume}{1},
  \bibinfo{number}{4} (\bibinfo{year}{1988}), \bibinfo{pages}{305--370}.
\newblock


\bibitem[\protect\citeauthoryear{Saltzer and Schroeder}{Saltzer and
  Schroeder}{1975}]%
        {Saltzer_Schroeder_75}
\bibfield{author}{\bibinfo{person}{Jerome~H. Saltzer} {and}
  \bibinfo{person}{Michael~D. Schroeder}.} \bibinfo{year}{1975}\natexlab{}.
\newblock \showarticletitle{The Protection of Information in Computer Systems}.
\newblock \bibinfo{journal}{\emph{Proc. IEEE}}  \bibinfo{volume}{63}
  (\bibinfo{year}{1975}), \bibinfo{pages}{1278--1308}.
\newblock


\bibitem[\protect\citeauthoryear{{seL4 Foundation}}{{seL4 Foundation}}{2021a}]%
        {sel4:perf}
\bibfield{author}{\bibinfo{person}{{seL4 Foundation}}.}
  \bibinfo{year}{2021}\natexlab{a}.
\newblock \bibinfo{title}{Performance}.
\newblock
\newblock
\urldef\tempurl%
\url{https://sel4.systems/About/Performance/}
\showURL{%
\tempurl}


\bibitem[\protect\citeauthoryear{{seL4 Foundation}}{{seL4 Foundation}}{2021b}]%
        {seL4:URL}
\bibfield{author}{\bibinfo{person}{{seL4 Foundation}}.}
  \bibinfo{year}{2021}\natexlab{b}.
\newblock \bibinfo{title}{The {seL4} Microkernel}.
\newblock
\newblock
\urldef\tempurl%
\url{https://sel4.systems/}
\showURL{%
\tempurl}


\bibitem[\protect\citeauthoryear{{seL4 Foundation}}{{seL4 Foundation}}{2023}]%
        {microkit:url}
\bibfield{author}{\bibinfo{person}{{seL4 Foundation}}.}
  \bibinfo{year}{2023}\natexlab{}.
\newblock \bibinfo{title}{{seL4} Microkit {GitHub}}.
\newblock
\newblock
\urldef\tempurl%
\url{https://github.com/seL4/microkit}
\showURL{%
\tempurl}


\bibitem[\protect\citeauthoryear{{seL4 Foundation}}{{seL4 Foundation}}{2024}]%
        {seL4:roadmap}
\bibfield{author}{\bibinfo{person}{{seL4 Foundation}}.}
  \bibinfo{year}{2024}\natexlab{}.
\newblock \bibinfo{title}{{seL4} Project Roadmap}.
\newblock
\newblock
\urldef\tempurl%
\url{https://docs.sel4.systems/projects/roadmap.html}
\showURL{%
\tempurl}


\bibitem[\protect\citeauthoryear{Sigurbjarnarson, Bornholt, Torlak, and
  Wang}{Sigurbjarnarson et~al\mbox{.}}{2016}]%
        {Sigurbjarnarson_BTW_16}
\bibfield{author}{\bibinfo{person}{Helgi Sigurbjarnarson},
  \bibinfo{person}{James Bornholt}, \bibinfo{person}{Emina Torlak}, {and}
  \bibinfo{person}{Xi Wang}.} \bibinfo{year}{2016}\natexlab{}.
\newblock \showarticletitle{Push-Button Verification of File Systems via Crash
  Refinement}. In \bibinfo{booktitle}{\emph{USENIX Symposium on Operating
  Systems Design and Implementation}}. \bibinfo{address}{Savannah, GA, US},
  \bibinfo{pages}{1--16}.
\newblock


\bibitem[\protect\citeauthoryear{Wang and Seltzer}{Wang and Seltzer}{2022}]%
        {Wang_Seltzer_22}
\bibfield{author}{\bibinfo{person}{Bingyao Wang} {and} \bibinfo{person}{Margo
  Seltzer}.} \bibinfo{year}{2022}\natexlab{}.
\newblock \showarticletitle{Tinkertoy: Build Your Own Operating Systems for
  {IoT} Devices}.
\newblock  \bibinfo{volume}{41}, \bibinfo{number}{11} (\bibinfo{year}{2022}),
  \bibinfo{pages}{4028--4039}.
\newblock


\bibitem[\protect\citeauthoryear{Welch}{Welch}{1991}]%
        {Welch_91}
\bibfield{author}{\bibinfo{person}{Brent Welch}.}
  \bibinfo{year}{1991}\natexlab{}.
\newblock \showarticletitle{The File System Belongs in the Kernel}. In
  \bibinfo{booktitle}{\emph{USENIX Mach Workshop}}.
\newblock


\bibitem[\protect\citeauthoryear{Wheeler}{Wheeler}{2001}]%
        {Wheeler_01}
\bibfield{author}{\bibinfo{person}{David~A. Wheeler}.}
  \bibinfo{year}{2001}\natexlab{}.
\newblock \bibinfo{title}{{SLOCCount}}.
\newblock \bibinfo{howpublished}{\url{http://www.dwheeler.com/sloccount/}}.
\newblock


\bibitem[\protect\citeauthoryear{Whitaker, Shaw, and Gribble}{Whitaker
  et~al\mbox{.}}{2002}]%
        {Whitaker_SG_02a}
\bibfield{author}{\bibinfo{person}{Andrew Whitaker}, \bibinfo{person}{Marianne
  Shaw}, {and} \bibinfo{person}{Steven~D. Gribble}.}
  \bibinfo{year}{2002}\natexlab{}.
\newblock \showarticletitle{Denali: A Scaleable Isolation Kernel}. In
  \bibinfo{booktitle}{\emph{SIGOPS European Workshop}}. \bibinfo{address}{St
  Emilion, FR}, \bibinfo{pages}{9--15}.
\newblock


\bibitem[\protect\citeauthoryear{Wienand and Macpherson}{Wienand and
  Macpherson}{2004}]%
        {Wienand_Macpherson_04}
\bibfield{author}{\bibinfo{person}{Ian Wienand} {and} \bibinfo{person}{Luke
  Macpherson}.} \bibinfo{year}{2004}\natexlab{}.
\newblock \showarticletitle{ipbench: A Framework for Distributed Network
  Benchmarking}. In \bibinfo{booktitle}{\emph{Conference for Unix, Linux and
  Open Source Professionals (AUUG)}}. \bibinfo{address}{Melbourne, Australia},
  \bibinfo{pages}{163--170}.
\newblock


\bibitem[\protect\citeauthoryear{Wikipedia}{Wikipedia}{2001}]%
        {KISS:wp}
\bibfield{author}{\bibinfo{person}{Wikipedia}.}
  \bibinfo{year}{2001}\natexlab{}.
\newblock \bibinfo{title}{{KISS} Principle}.
\newblock
\newblock
\urldef\tempurl%
\url{https://en.wikipedia.org/wiki/KISS_principle}
\showURL{%
\tempurl}


\bibitem[\protect\citeauthoryear{Zaostrovnykh, Pirelli, Iyer, Rizzo, Argyraki,
  and Candea}{Zaostrovnykh et~al\mbox{.}}{2019}]%
        {Zaostrovnykh_PIRPAC_19}
\bibfield{author}{\bibinfo{person}{Arseniy Zaostrovnykh},
  \bibinfo{person}{Solal Pirelli}, \bibinfo{person}{Rishabh Iyer},
  \bibinfo{person}{Matteo Rizzo}, \bibinfo{person}{Luis Pedrosa~Katerina
  Argyraki}, {and} \bibinfo{person}{George Candea}.}
  \bibinfo{year}{2019}\natexlab{}.
\newblock \showarticletitle{Verifying Software Network Functions with No
  Verification Expertise}. In \bibinfo{booktitle}{\emph{ACM Symposium on
  Operating Systems Principles}}.
\newblock


\bibitem[\protect\citeauthoryear{Zaostrovnykh, Pirelli, Pedrosa, Argyraki, and
  Candea}{Zaostrovnykh et~al\mbox{.}}{2017}]%
        {Zaostrovnykh_PPAC_17}
\bibfield{author}{\bibinfo{person}{Arseniy Zaostrovnykh},
  \bibinfo{person}{Solal Pirelli}, \bibinfo{person}{Luis Pedrosa},
  \bibinfo{person}{Katerina Argyraki}, {and} \bibinfo{person}{George Candea}.}
  \bibinfo{year}{2017}\natexlab{}.
\newblock \showarticletitle{A Formally Verified {NAT}}. In
  \bibinfo{booktitle}{\emph{ACM Conference on Communications}}.
\newblock


\bibitem[\protect\citeauthoryear{Zhao, Legnani, Ung, Truong, Sau, Tanaka,
  Pohjola, Sewell, Sison, Syeda, Myreen, Norrish, and Heiser}{Zhao
  et~al\mbox{.}}{2025}]%
        {Zhao_LTUTSTASSSMNH_25}
\bibfield{author}{\bibinfo{person}{Junming Zhao}, \bibinfo{person}{Alessandro
  Legnani}, \bibinfo{person}{Tiana~Tsang Ung}, \bibinfo{person}{H. Truong},
  \bibinfo{person}{Tsun~Wang Sau}, \bibinfo{person}{Miki Tanaka},
  \bibinfo{person}{Johannes~{\AA}man Pohjola}, \bibinfo{person}{Thomas Sewell},
  \bibinfo{person}{Rob Sison}, \bibinfo{person}{Hira Syeda},
  \bibinfo{person}{Magnus Myreen}, \bibinfo{person}{Michael Norrish}, {and}
  \bibinfo{person}{Gernot Heiser}.} \bibinfo{year}{2025}\natexlab{}.
\newblock \showarticletitle{Verifying Device Drivers with {Pancake}}.
\newblock \bibinfo{journal}{\emph{arXiv preprint}} (\bibinfo{date}{Jan.}
  \bibinfo{year}{2025}), 15.
\newblock
\urldef\tempurl%
\url{https://arxiv.org/abs/2501.08249}
\showURL{%
\tempurl}


\end{thebibliography}
}
\end{document}